\documentclass[letterpaper,11pt]{article}

\usepackage[utf8]{inputenc}
\usepackage[english]{babel}
\usepackage{mathpazo}              
\usepackage[scaled=0.92]{helvet}
\usepackage{csquotes}              

\usepackage{amsmath, amssymb, amsfonts, amsthm}
\usepackage{mathtools}
\usepackage{bbm}
\usepackage{comment}

\theoremstyle{remark}

\usepackage{enumitem}
\setlist[description]{font=\normalfont\itshape}

\usepackage{graphicx}
\usepackage{color}
\usepackage{tabularx}
\usepackage{adjustbox}
\usepackage{ragged2e}
\usepackage{booktabs}
\usepackage{longtable}
\usepackage{threeparttable}        
\usepackage{caption}
\usepackage{lscape}
\usepackage{pifont}

\usepackage{float}
\usepackage{xcolor}

\newcolumntype{L}[1]{>{\raggedright\arraybackslash}p{#1}}
\newcolumntype{C}[1]{>{\centering\arraybackslash}p{#1}}
\newcolumntype{R}[1]{>{\raggedleft\arraybackslash}p{#1}}

\definecolor{linkgray}{RGB}{80,100,150}

\usepackage[%
    authordate,
    backend=biber,
    eprint = false,
    doi = false,
    maxcitenames=2,
    natbib,
    noibid,
    footmarkoff
    ]{biblatex-chicago}

\DeclareCiteCommand{\citeyearpar}
    {}
    {\mkbibparens{\bibhyperref{\printdate}}}
    {, }
    {}

\newbibmacro{string+url}[1]{%
\iffieldundef{url}{#1}{\href{\thefield{url}}{#1}}
}

\DeclareFieldFormat{title}{\usebibmacro{string+url}{\mkbibemph{#1}}}
\DeclareFieldFormat[article,incollection]{title}%
    {\usebibmacro{string+url}{\mkbibquote{#1}}}

\DeclareFieldFormat{citehyperref}{%
  \DeclareFieldAlias{bibhyperref}{noformat}
  \bibhyperref{#1}}

\DeclareFieldFormat{apacase}{#1} 

\DeclareFieldFormat{textcitehyperref}{%
  \DeclareFieldAlias{bibhyperref}{noformat}
  \bibhyperref{%
    #1%
    \ifbool{cbx:parens}
      {\bibcloseparen\global\boolfalse{cbx:parens}}
      {}}}

\savebibmacro{cite}
\savebibmacro{textcite}
\renewbibmacro*{cite}{%
  \printtext[citehyperref]{%
    \restorebibmacro{cite}%
    \usebibmacro{cite}}}
\renewbibmacro*{textcite}{%
  \ifboolexpr{
    ( not test {\iffieldundef{prenote}} and
      test {\ifnumequal{\value{citecount}}{1}} )
    or
    ( not test {\iffieldundef{postnote}} and
      test {\ifnumequal{\value{citecount}}{\value{citetotal}}} )
  }
    {\DeclareFieldAlias{textcitehyperref}{noformat}}
    {}%
  \printtext[textcitehyperref]{%
    \restorebibmacro{textcite}%
    \usebibmacro{textcite}}}

\usepackage{url}
\usepackage{epigraph}
\usepackage{csvsimple}
\usepackage{setspace}
\usepackage[left=2.5cm,top=2.5cm,right=2.5cm,bottom=2.5cm]{geometry}
\usepackage{listings}
\usepackage[hidelinks]{hyperref}

\begin{document}

\begin{titlepage}

\renewcommand{\thefootnote}{\fnsymbol{footnote}} 
\vspace*{10 mm}

\begin{center}
{\Large 
Hysteresis and Selection in the Rise of Fascism: \\
\medskip
The `Ordinary Men' of the Nazi Party}
 \\[8 mm]
Luis Bosshart\\
Max Deter\\
Leander Heldring\\
Cathrin Mohr\\
Matthias Weigand\footnote[1]{Bosshart: Harvard Academy for International and Area Studies. Email: \texttt{lbosshart@fas.harvard.edu}. Deter: Potsdam University, Department of Economics, and Berlin School of Economics. Email: \texttt{max.deter@gmail.com}. Heldring: Northwestern University, Kellogg School of Management. Email: \texttt{leander.heldring@kellogg.northwestern.edu}. Mohr: Kiel Institute for the World Economy \& University of Hamburg. Email: \texttt{cathrin.mohr@kielinstitut.de}. Weigand: Harvard University, Department of Economics. Email: \texttt{mweigand@g.harvard.edu}. We thank Ed Glaeser for helpful comments and Zach Kaplan for generous technical expertise. We thank the German Federal Archives (\textit{Bundesarchiv}) for ongoing discussions and for generously sharing expertise on the archival materials throughout 2025. We thank Elena Damianakis and Ekaterina Kulebiakina for excellent research assistance. Financial support from the Kenneth C. Griffin Economics Research Fund is gratefully acknowledged. 
} 
\\[0.5cm] 
\textit{First version: {\color{blue}\href{https://arxiv.org/abs/2604.17697}{April 20, 2026}}}\\
\textit{Latest version: \today}

\vspace*{3 mm}

\end{center}

\vspace*{0.2 cm}
\begin{abstract}
\noindent

We digitize and analyze the near-universe of National Socialist German Workers' Party (NSDAP) membership records and link them to population and industrial censuses. Four findings emerge. First, as the party expanded, its membership came to resemble the broader population more closely in occupational, demographic, and religious terms. Second, SS members' characteristics remained different: younger, more educated, and more fanatical, as measured by the display of Nazi insignia in membership portraits. Third, within communities, coworkers, and families, early membership generated hysteresis, with subsequent entrants drawn from the same groups. Finally, local increases in party membership are associated with subsequent deportations of Germany's Jews.
\end{abstract}

 { \small
 \begin{quote}
 \end{quote}
 }

\end{titlepage}

\renewcommand{\thefootnote}{\arabic{footnote}} 

\setlength{\epigraphwidth}{0.6\textwidth}

\epigraph{\textit{At the same time, however, the collective behavior of Reserve Police Battalion 101 has deeply disturbing implications. There are many societies afflicted by traditions of racism and caught in the siege mentality of war or threat of war. \dots Within virtually every social collective, the peer group exerts tremendous pressures on behavior and sets moral norms. If the men of Reserve Police Battalion 101 could become killers under such circumstances, what group of men cannot?}}{Christopher R.\ Browning, \textit{Ordinary Men} (1992)}
\doublespacing

\section{Introduction}

Whether members of the Nazi Party (NSDAP) were ``ordinary men'' or ideological fanatics has been a central question across the historical, psychological, and social sciences. Hannah Arendt famously advanced the thesis of the ``banality of evil,'' arguing that perpetrators of mass atrocity could be mundane bureaucrats rather than ideological zealots \citep{arendt1963eichmann}. Psychologists have reinforced this interpretation through experimental evidence on obedience to authority and conformity \citep{milgram1963,milgram1974,zimbardo2007}. Historians and political scientists, meanwhile, have examined the social composition of the NSDAP and its paramilitary arm, the SS, asking whether members were drawn selectively from particular social groups or were broadly representative of German society \citep{browning1992ordinary,goldhagen1996willing,muhlberger1991,mann2004}. In Germany, the debate has also been sociological: if perpetrators were not socially distinctive, then the conditions under which similar movements can mobilize masses may be general \citep{waller2002}. These questions are largely unresolved---scholars continue to contest, for instance, whether Adolf Eichmann was a committed antisemite \citep{arendt1963eichmann,cesarani2004,stangneth2014}.

In this paper, we assemble the near-universe of NSDAP and SS membership records, expanding the record base from roughly 50{,}000 individuals to approximately 11 million. We digitize and analyze individual NSDAP membership cards, extracting each member's name, birth date, occupation, residence, and dates of party entry and exit. We link these records to newly digitized German population and industrial censuses. We use these data to examine differences between members and non-members. In the spirit of \citet{browning1992ordinary}, we study the demographic, residential, and occupational selection of NSDAP and SS members. Unlike the prior literature, our individual-level data allow us to trace social and demographic differences over space and time in unprecedented detail.\footnote{Compared with prior work, our data improve spatial resolution by more than a factor of 50 and allow us to trace membership dynamics at the monthly level, including around key events such as the Nazi seizure of power in 1933.} Like the existing literature, our approach cannot directly measure underlying psychological or attitudinal differences.

We start by documenting the patterns and dynamics of NSDAP membership. We find that entry into the NSDAP occurred in sharp, discontinuous waves, most prominently in 1933 following the seizure of power and in 1937, after a membership ban was lifted. Membership also varied substantially across space, with far greater variation within counties than between them. We find that nearly all spatial variation arises \textit{within} counties, the finest administrative unit available in prior research. Moreover, 40\% of municipalities recorded no entry at all.

We then study the characteristics of joiners, distinguishing between regular party members and members of the paramilitary SS. Early NSDAP entrants differed from the general population: they were disproportionately middle class and overwhelmingly male. They do not appear to have been more likely to be Protestant. These differences narrowed over time, and by the outbreak of the Second World War the Nazi Party had become a mass movement. The SS remained more selected: its members were younger, better educated, and more likely to come from professional occupations. Relative to NSDAP members, they also appear to have been more ideologically committed. We trace this using the display of National Socialist insignia in membership portraits taken at entry.

These patterns motivate the second part of the paper, in which we study dynamics of mass entry. We first examine how early joiners shaped the wave-like and spatially concentrated pattern of membership. We find a ``hysteresis'' pattern: within municipalities, municipality-by-occupation, and municipality-by-family cells, early members were followed by later members from the same groups. We then document stronger acceleration during the March 1933 surge in municipalities where the preexisting party base included more women, a paramilitary presence, and groups central to the diffusion of local norms, including doctors, policemen, and especially teachers.

Finally, we relate membership to the Holocaust. In municipality-years that feature more NSDAP entry, we observe higher emigrations and deportations.

Our paper makes three contributions. First, we provide a new dataset: the near-universe of NSDAP membership records, linked to census data, enabling individual-level analysis of a phenomenon previously studied only in aggregate. A simple variance decomposition demonstrates why this resolution matters: around 95\% of the total variation in municipal NSDAP membership shares arises \textit{within} counties, not between them. The existing literature, constrained to county-level data, has by construction studied only the smaller, between-county component. Second, we document facts that are invisible in aggregate data: the ordinariness of joiners, the distinctiveness of the SS, and the wave-like dynamics of entry. Together, these shift the locus of explanation from individual selection toward social coordination.  Third, we show that party growth was highly localized and persistent within communities, occupations, and families, and that local party strength is positively associated with subsequent anti-Jewish persecution.

The question of how the Nazi Party rose to power has generated interest like few others in the social sciences. We contribute to the vast literature spanning history, political science, and economics that investigates the origins of Nazism.

One influential line of research emphasizes Germany's distinctive historical trajectory. This is often summarized under the \textit{Sonderweg} thesis, according to which Germany followed a historically distinctive path that increased the likelihood of authoritarian outcomes \citep{steinmetz1996german}.  Within this view, some scholars emphasize persistent antisemitism and elements of political culture that made parts of the German population receptive to fascism \citep[for example][]{pulzer1964rise,goldhagen1996willing,evans2003coming}. A second account holds that Germany's path to modernity was distinctive in ways that made democratic consolidation fragile \citep[for example][]{rosenberg1958bureaucracy,kocka1980ursachen,moore1966social}.  An alternative view stresses that Nazism reflected broader crises of industrial modernity rather than uniquely German predispositions, consistent with the fact that nationalist and antisemitic movements gained ground across interwar Europe \citep{evans2003coming}. In this perspective, mass authoritarian movements can arise under general conditions through coordination rather than deep-rooted ideological commitment \citep[e.g.][]{blackbourn1984peculiarities,peukert1992weimar}.

Within this broader debate, historians have examined whether participants were ideological fanatics or ordinary men swept up by circumstance \citep{browning1992ordinary, goldhagen1996willing, arendt1951origins, boehnert1977ss}. Political scientists have shown that the NSDAP drew disproportionately from Protestant, rural, and middle-class constituencies \citep{falter1991hitlers, childers1983nazi}, with \citet{kater1983nazi} providing the foundational social profile of party members and \citet{falter2020parteigenossen} offering the first large-scale quantitative analysis based on approximately 50,000 records. In economics, research has emphasized the role of economic shocks \citep{king2008ordinary, doerr2022financial, galofrevila2021austerity}, cultural persistence \citep{voigtlander2012persecution, becker2023death}, media and propaganda \citep{adena2015radio, caesmann2025viral, voigtlander2026highway}, and social capital \citep{satyanath2017bowling, ferguson2008betting}. 

A common limitation across these contributions is their reliance on aggregate (voting) data or small, potentially non-random samples of members. Our data overcome this constraint by providing the near universe of NSDAP and SS membership records at the individual level. This allows us to test existing hypotheses on the origins of Nazism at the level of individual joining decisions while accounting for local context. In addition, by jointly analyzing mass party membership and selective elite recruitment into the SS, we shed light on the distinct roles of broad-based mobilization and ideological selection within the Nazi movement.

Our paper also relates to the economics of political change and conflict participation. \citet{sanchezdelasierra2025enlistment} studies self-selection into violence, finding that joiners exhibit parochial altruism \citep{bowles2011cooperative}. \citet{viterna2006pulled} traces heterogeneity in pathways into insurgent mobilization. We document a complementary mechanism: coordination dynamics in which the decision to join depends on the decisions of others. More broadly, our analysis provides micro-level evidence on the coordination process through which citizens come to support a new regime, a mechanism central to models of political change \citep{acemoglu2006economic} but rarely observed empirically.

The rest of the paper proceeds as follows. Section~\ref{sec:background} provides institutional and historical context on the rise of the NSDAP and the organization of the party. Section~\ref{sec:data} describes the construction of the individual-level dataset and the additional sources used in the analysis. Section~\ref{sec:descriptives} presents descriptive evidence on the composition and geographic distribution of party membership. Section~\ref{sec:coordination} examines coordination dynamics in municipal entry. Section~\ref{sec:consequences} examines the consequences of local party strength. The Appendix provides additional details on the data, measurement, and robustness analyses.

\clearpage
\section{Institutional and Historical Context}
\label{sec:background}
We study the rise of fascism by analyzing the National Socialist German Workers' Party (NSDAP). The party grew from a fringe movement of a few thousand members in 1920 to a mass organization of roughly 11 million by 1945, encompassing about one in six German adults. It is one of the most consequential political movements in history, transforming institutions and organizing unprecedented state violence. 

This rise took place in a broader interwar context of political instability, ideological polarization, and intense competition between liberal, socialist, and nationalist projects \citep{nolte1965three}.

\begin{figure}[!t]
\caption{NSDAP Membership Over Time}
\centering
\vspace{-.4cm}
\includegraphics[width=\textwidth]{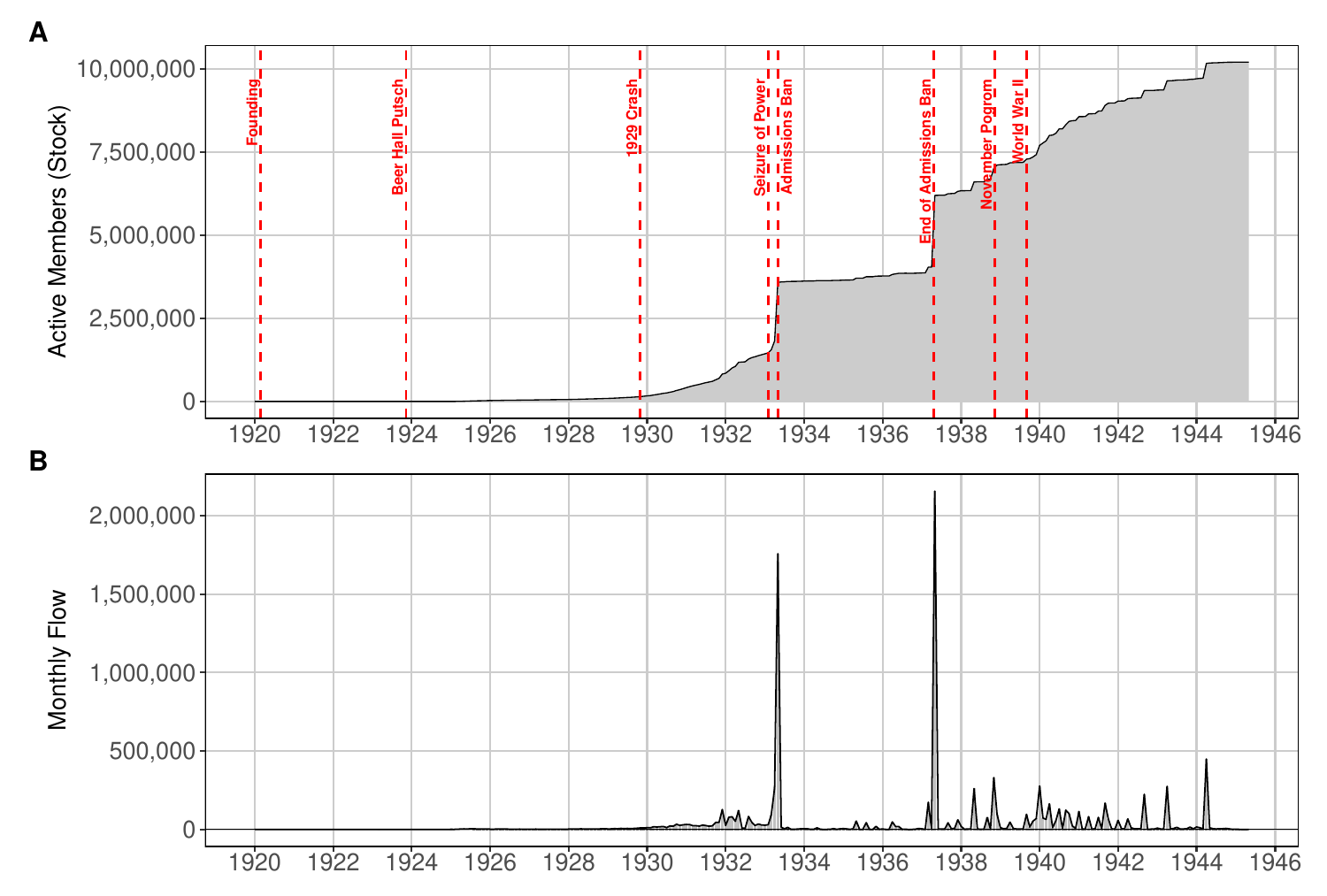}
\label{fig:membership_time}
\\[4pt]
\justifying
\vspace{-.75cm}
\scriptsize{\textbf{Note.} Panel~A shows cumulative active membership stock by month. Panel~B shows monthly joiners. Dashed red lines mark key events: party founding (1920), Beer Hall Putsch (1923), Wall Street Crash (1929), seizure of power (1933), Night of the Long Knives (1934), the lifting of the membership freeze (1937), Kristallnacht (1938), and World War~II (1939).} 
\end{figure}

The NSDAP was founded in Munich on February~24, 1920, as the successor to the far-right German Workers' Party (DAP). Figure~\ref{fig:membership_time} plots the evolution of party membership, showing both the stock of members and the flow of new entrants over time.\footnote{Because not all membership cards survived, our data provide a lower bound on total party entry. We discuss patterns of missingness in Section~\ref{sec:data}.}

The NSDAP formed part of a broader nationalist, antisemitic, and authoritarian movement in Germany. Following the onset of the Great Depression in 1929, it transformed from a marginal party into a mass movement. While the party received less than 3 percent of the vote in the Reichstag election of May 1928, its vote share rose to 18.3 percent in September 1930 and reached 37.3 percent by July 1932. This political rise was also accompanied by a massive increase in party membership, which expanded to approximately 850{,}000 by late~1932 \citep{kater1983nazi}. 

Given the instability of the Weimar Republic, there was considerable uncertainty about which political equilibrium would ultimately prevail. Under those conditions, joining the NSDAP still functioned as a signal of ideological commitment \citep{bracher1969deutsche, bracher1955auflosung}. That uncertainty persisted well into late 1932, when the party was in visible decline: it lost roughly two million votes across two elections, and its chief organizer, Gregor Strasser, defected.

The first months of 1933 marked the decisive break. The Lippe state election of January 15, in which the NSDAP emerged as the largest party, offered an early indication of reversal. Hitler’s appointment as chancellor on January 30 was followed by the Reichstag fire of February 27, which furnished the pretext for the arrest of Communist leaders and the overnight suppression of the opposition. In the Reichstag election of March 5, the NSDAP won 44 percent of the vote. Within a week, the regime had installed commissars across the German states \citep{bracher1969deutsche}. Goebbels observed in late February that police officers who had previously been hostile were ``now all Nazis,'' \citep[entry of 24~February 1933, vol.~2/III, p.~134]{goebbels2006} and in early March wrote that ``resistance was futile'' \citep[entry of 9~March 1933, vol.~2/III, p.~143]{goebbels2006}. No further multiparty Reichstag elections were held, and the NSDAP remained in power until Germany’s defeat in World War II.

Our data suggest that uncertainty about the regime was resolved late: the membership surge in Figure~\ref{fig:membership_time} appears only after the March election. Civil servants and teachers were especially prominent among these new entrants \citep{childers1983nazi}.


In addition to individual motivations for joining the party, membership was shaped by party-specific and local characteristics. Two features of the NSDAP were particularly important. First, although the party was organized as a hierarchical \textit{Führerpartei}, recruitment and implementation were decentralized at the local level.%
\footnote{The NSDAP was organized into local branches (\textit{Ortsgruppen}) grouped into regional districts (\textit{Gaue}), each headed by a \textit{Gauleiter} appointed by the party leadership \citep{bracher1969deutsche, evans2003coming}. Despite this formal hierarchy, the party was institutionally fragmented, and regional leaders competed for influence with other powers in the party, with authority flowing through personal loyalty to Hitler rather than bureaucratic chains of command \citep{broszat1981hitler, mommsen1991nationalsozialismus}. Hitler imposed the \textit{F\"{u}hrerprinzip} at every level, eliminating many formal decision-making bodies \citep{mommsen1991nationalsozialismus}. This meant that \textit{Gauleiter} retained pronounced autonomy, as they could only be displaced by Hitler himself, not by any party bureaucracy \citep{broszat1981hitler}.}
At the municipal level, local \textit{Ortsgruppenleiter} controlled admission to the party. Because local leaders exercised substantial discretion over recruitment and enforcement, the conditions shaping individual membership decisions varied sharply across communities.

Second, after a surge in applications, the NSDAP imposed a membership freeze (\textit{Mitgliedersperre}) on May~1, 1933, ostensibly to prevent purely opportunistic entry. The freeze itself signaled the leadership's recognition that entry had shifted from ideological commitment to mass, strategic joining.

In addition, regional party penetration depended on local social structure, associational density, the strength of political competition and whether early committed activists served as coordination nodes \citep{allen1984nazi}. 


\textbf{The Schutzstaffel (SS)}

While NSDAP membership broadened into a mass phenomenon, the regime also relied on more selective organizations. The \textit{SS}, founded in 1925 as a small personal guard for Hitler, over time expanded into an elite organization with its own command structure, intelligence apparatus, and screening process: admission required proof of racial purity, physical fitness, and, for higher ranks, demonstrated educational attainment \citep{boehnert1977ss,longerich2012heinrich}. While the NSDAP coordinated the masses, the SS selected for ideological commitment and specific personal characteristics. This contrast between broad party mobilization and selective elite recruitment is central to our analysis.

\begin{figure}[!t]
\centering
\caption{Membership Exits and SS Membership}
\vspace{-0.3cm}
\includegraphics[width=\textwidth]{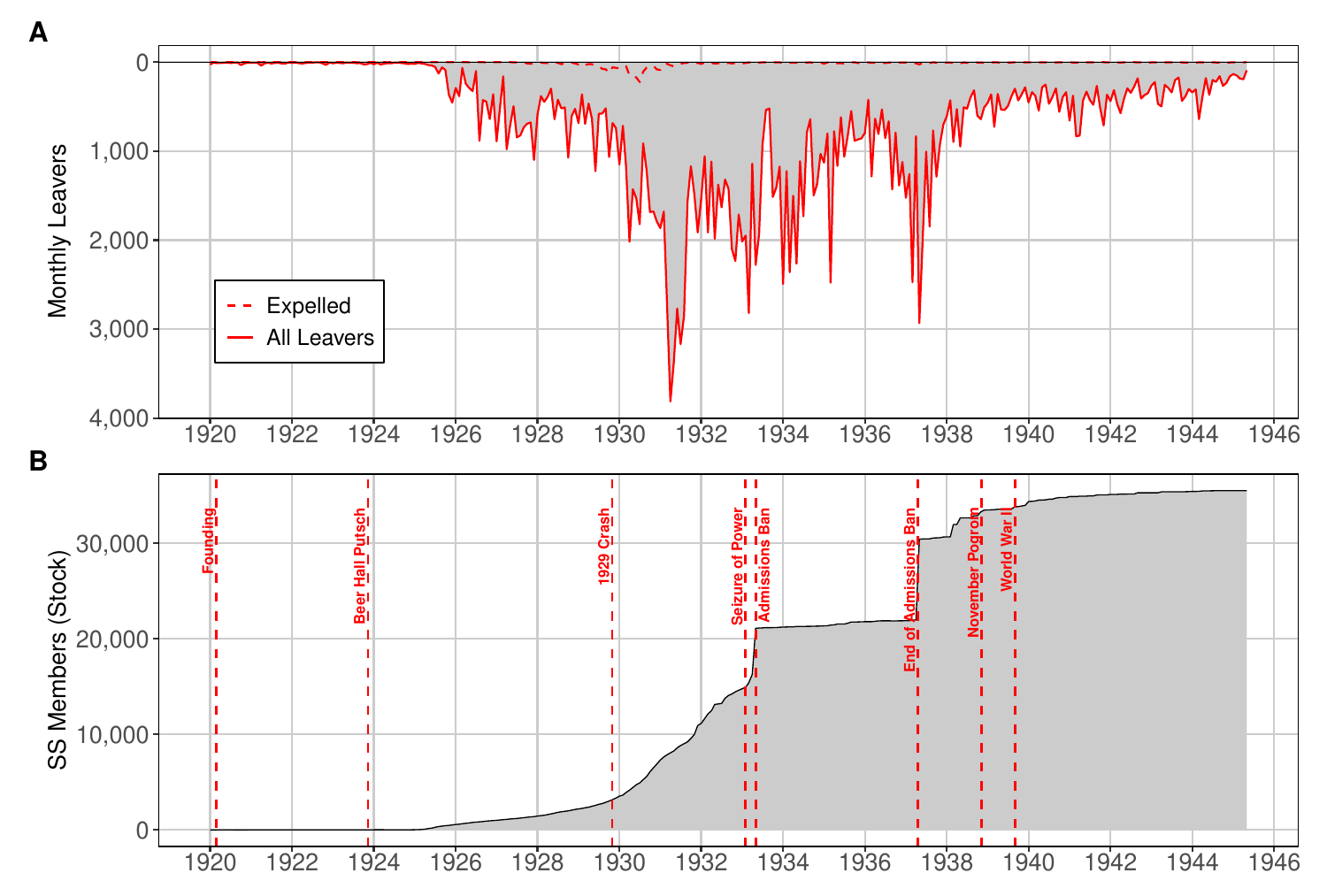}
\label{fig:membership_leavers_ss}
\vspace{4pt}
\begin{minipage}{\textwidth}
\scriptsize
\vspace{-1cm}
\textbf{Note.} Panel~A: Grey area shows the monthly flow of all membership exits. Red solid line shows all leavers; red dashed line shows expelled members only. Panel~B: NSDAP entry years for eventual SS members. Dashed red vertical lines mark key historical events.
\end{minipage}
\end{figure}

Figure~\ref{fig:membership_leavers_ss}, Panel~B examines NSDAP entry years for eventual SS members. A large share of this group also entered during the two main waves in 1933 and 1937, indicating that even the pool from which the SS was drawn was shaped by the same aggregate entry dynamics.

Who joined the party matters as much as when and at what scale. In pioneering work based on a 0.5\% sample of members, \citet{falter1991hitlers}, \citet{kater1983nazi}, and \citet{childers1983nazi} find that the party attracted a cross-class coalition: the lower middle class (artisans, shopkeepers, and independent professionals) was overrepresented, but large numbers of workers, farmers, and civil servants also entered the party, and white-collar employees (\textit{Angestellte}) were particularly prominent \citep{kater1983nazi,kocka1980ursachen}. Prior aggregate statistics suggest that membership was skewed toward Protestant, rural, and small-town constituencies.

Qualitative evidence suggests that motivations to join the party were heterogeneous and changed over time: antisemitism, anti-Marxism, national renewal, economic insecurity, career considerations, and youthful idealism all appear in contemporary accounts \citep{abel1938why, merkl1975political}.

\begin{figure}[!t]
\caption{Motives for Joining in Abel Autobiographies}
\centering
\vspace{-.25cm}
\includegraphics[width=0.75\textwidth]{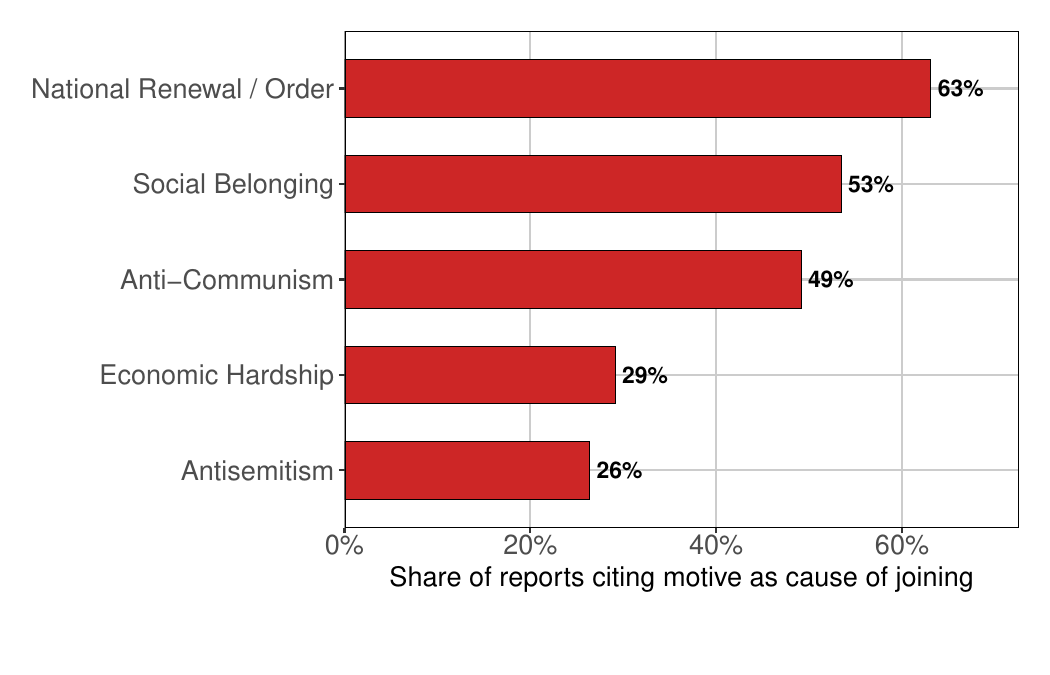}
\label{fig:abel_motives}
\\[4pt]
\vspace{-.75cm}
\justifying
\scriptsize{\textbf{Note.} Share of $N = 580$ Abel autobiographies in which each motive is coded as an explicit causal reason for joining the NSDAP. A report can be coded for multiple motives. Coding via structured-prompt LLM classification (GPT-4o-mini); reports shorter than 80 words are excluded. Appendix~\ref{sec:appendix_abel} describes the coding procedure and reports example passages.}
\end{figure}

Figure~\ref{fig:abel_motives} quantifies this heterogeneity. We code 580 first-person accounts written in 1934 by party members invited to explain why they joined \citep[see][]{abel1938why}, classifying each into five categories of motive: national renewal and order, social belonging, anti-communism, economic hardship, and antisemitism. We record a motive only when the text explicitly links it to the decision to join; background characteristics, general ideological statements, and post-entry activity are excluded. Appendix~\ref{sec:appendix_abel} details the coding procedure and reports example passages. National renewal appears in 63 percent of reports, social belonging in 53 percent, and anti-communism in 49 percent; economic hardship (29 percent) and antisemitism (26 percent) are cited as causes less often. Motives overlap within reports: 69 percent of accounts cite two or more, with an average of 2.2 per report, consistent with multiple concurrent drivers of entry rather than a single dominant cause. The Abel sample is self-selected as they replied to a prize competition and skews toward early, committed joiners, so we treat these shares as qualitative corroboration of the heterogeneity documented in contemporary accounts rather than as representative of the full membership. We return to some of these motives, in particular the social dynamics of entry, suggestively in Section~\ref{sec:coordination}.

\section{Data}
\label{sec:data}

\subsection{Individual-Level Data}
\label{sec:data_individual}

We construct an individual-level dataset of NSDAP membership from the near-universe of surviving membership cards held by the U.S. National Archives (NARA, Record Group~242). The collection comprises approximately 16.3 million card images, corresponding to roughly 11~million distinct members, across two parallel indexes: the \textit{Zentralkartei} (alphabetical, central index) and the \textit{Ortsgruppenkartei} (geographic, by local branch). According to \citet{falter2020parteigenossen}, over 90 percent of all persons who ever joined the NSDAP are traceable across the two indexes. We describe the extraction pipeline in Appendix~\ref{app:data_individual}.\footnote{Allied forces captured these records in 1945 and consolidated them at the Berlin Document Center, where they were microfilmed. The \textit{Zentralkartei} consists of  approximately 9.5 million images, organized alphabetically by surname, and the \textit{Ortsgruppenkartei} of approximately 6.7 million images, organized geographically by \textit{Gau}. Many members appear in only one index. Records for roughly one million members have been lost: the \textit{Zentralkartei} has gaps in the letter ranges Fa--G and Ka--O; the \textit{Ortsgruppenkartei} lost major holdings from Vienna, Salzburg, and Franconia. The majority of these missing cards, thus, pertain to areas outside of current-day Germany.}

We extract structured data from each card image using a vision-language model, which returns over 40 fields per card, including: name, birth date, place of birth, occupation, membership number, entry date, residential address, local branch, regional district, and administrative annotations recording exits, expulsions, re-entries, and deaths. Other fields typically include repetitions of these fields.

The occupation field records free-text entries with extensive German abbreviations. We standardize these entries and classify each into the occupational categories of the 1925 census (\textit{Berufszählung}), following the Statistisches Reichsamt classification (Appendix~\ref{sec:occupation}). We also retain alternative occupation codings based on ISCO-08, NACE Rev.~2, and the census professional groups (\textit{Berufsgruppen}).\footnote{Self-reported occupations may differ from official census records due to social desirability or multiple job-holding. For example, agricultural workers frequently held secondary jobs, which may explain the differences between party and census data in agriculture.} To assign individuals to local workplace, we combine data on occupational status with data on industry. In the main specification, workplace cells are defined at the municipality-by-industry-by-occupational-category level. We further map occupations to a hierarchy of state functions following \citet{boberach2012amtsbezeichnungen}, identifying members in public administration, police, military, or party organizations.

Extracting structured data from historical handwritten documents introduces measurement error that may not be classical. Legibility varies with card age, regional handwriting conventions, and the care of the local official who completed the form. Two card formats coexist: small index cards with core biographical data, and larger A5 cards with address histories and administrative annotations. In our initial sample, approximately 2--5\% of fields per card contain the model's \texttt{[unclear]} annotation. We cannot fully rule out that extraction accuracy varies systematically with card characteristics that correlate with the outcomes we study. Our pipeline takes into account that cards vary in format: some are machine-printed with standardized fields, while others were completed in cursive by local officials (Appendix Figures~\ref{fig:card_printed} and~\ref{fig:card_handwritten}). Extraction is more accurate for machine-printed cards. We examine this variation in robustness checks below.

We validate the extracted data internally. Manual transcription of a random sample of 1,000 cards yields 92.1\% field-level agreement. Fill rates for core analytic variables (name, birth date, address, local branch, membership number) range from 83\% to 100\%. The occupation field has a fill rate of 87\%, reflecting variation in card completeness rather than systematic extraction failure.

To assess whether non-classical measurement error biases our estimates, we conduct two robustness exercises. First, we restrict the sample to machine-printed cards, which are more legible and less susceptible to extraction error. All main results are unchanged. Second, we control for card-level proxies of extraction difficulty: an indicator for handwritten cards and a word-count measure of layout complexity. If measurement error were driving our results, conditioning on these proxies would attenuate the estimates. It does not. Together, these exercises suggest that our findings are not an artifact of differential extraction quality.

We match extracted records to historical municipalities using the \textit{Meyers Orts- und Verkehrslexikon des Deutschen Reichs}, a geocoded gazetteer of German place names. We construct municipal boundaries through population-weighted Voronoi tessellation within county borders, yielding approximately 65,000 municipal polygons in the 1910 \textit{Gemeindeverzeichnis} (Appendix~\ref{sec:municipal_boundaries}). This spatial linkage allows us to compute municipal-level membership shares by dividing NSDAP member counts, identified by address, by (imputed) census population at the municipal level to get disaggregated population shares.

We identify SS membership from two sources. First, we flag all members whose NSDAP card lists an ``SS'' affiliation. Second, we link our records to approximately 50,000 individuals named in the \textit{SS-Verordnungsblatt}, the SS's official gazette \citep{dwsxip_ss_verordnungsblatt}, matching on membership number where available and on name and date of birth otherwise. The external source provides SS entry date, rank, and promotion history. In total, we conservatively link over 46,000 individuals in the SS to the NSDAP records.

We estimate each member's probability of being Protestant using a Bayesian framework that combines county-level Protestant shares from the 1925 census, municipal-level church counts from the \textit{Meyers Orts- und Verkehrslexikon}, and name-specific likelihood ratios estimated from 1.6 million parish register records \citep{verein2023ofb}. Appendix~\ref{sec:appendix_protestant} provides details and reports the full name-religion association table. We validate this method with a hold-out sample and obtain 0.98 accuracy.

We construct two within-municipality social proximity measures from the address and occupation data. For each member and year, we count existing NSDAP members who share the same surname (family), or workplace cell (coworker). The full set of extracted fields and their fill rates is reported in Appendix Table~\ref{tab:variables}.

The resulting dataset operates at three levels: individual (member characteristics, entry and exit timing), municipal (annual panels of membership stocks and flows), and county (membership aggregates merged with census and administrative data described below). Appendix Table~\ref{table:summary} reports summary statistics.

\subsection{Cross-Section Data}
\label{sec:data_cross}

We supplement the individual-level panel with a set of variables that capture baseline county- and location-level conditions.

\paragraph{Physical Geography.}

At the county level, we measure area, city density, mean elevation, terrain ruggedness \citep{eea2013eudem}, and agricultural suitability \citep{fao2002gaez}. At the location level, we compute distances to the coastline and to the nearest navigable river \citep{kunz99}. We also record latitude and longitude.

\paragraph{Economic Geography.}


We digitize the 1925 German industrial census to obtain county-level employment counts by industry. We draw on the German population census of 1925 to construct population denominators and obtain the demographic and occupational composition of the general population at the county level. We use the occupational distributions to compare NSDAP and SS members with the general population.

\paragraph{Administrative Geography.}

For each county, we record its governing district and state. We construct indicators for whether the county belonged to Prussia, the largest state in Imperial Germany, and whether it formed part of the North German Confederation, the political alliance preceding unification. For each location, we compute distances to district, state, and national borders, as well as distances to the respective district, state, and national capitals.



%

\paragraph{World War II: Deportations.}

We obtain individual-level records on Jewish residents, deportees, and emigrants in the List of Jewish Residents in Germany 1933-1945, the 1939 German Minority Census and the German Federal Archives' memorial book \citep{tracingthepast2018mappingthelives, bundesarchiv2012gedenkbuch}. The database contains approximately 406,000 entries with residential addresses, deportation dates, and emigration records. We geocode each record to a municipality and construct municipal-level counts of Jewish residents, deportees, and emigrants, which serve as outcome variables in Section~\ref{sec:consequences}. From the same victim records, we construct a measure of Jewish presence at the address level for each NSDAP member. We parse victim residential addresses into street name and house number, apply the same street-name normalization used for the membership cards, and match to party members at three levels of proximity: same municipality, same street, and same house number.

\paragraph{Jewish Communities.}

We compile historical Jewish community population counts from approximately 4,600 localities across Germany, drawing on the \textit{J\"{u}dische Gemeinden} online archive \citep{juedischegemeinden}. We match locality names to our municipality coordinates and aggregate to the municipal level. These counts serve as a control variable for the local Jewish population share in the cross-sectional analysis.

\paragraph{Political Data.}
We collect county-level Weimar-era election results, including vote shares for the major parties in the Reichstag elections of 1928, 1930, July~1932, November~1932, and March~1933. Of particular interest are the KPD (Communist) and SPD (Social Democratic) vote shares, which measure local exposure to left-wing mobilization.

\section{Descriptive Evidence}
\label{sec:descriptives}
We document six facts about NSDAP and SS membership. The first two concern temporal dynamics and geographic structure: entry occurred in discrete waves, and most variation in membership rates is within counties. The next two characterize composition: joiners converge toward the general population over time, and members originate from representative counties. The final two examine individual-level differences: joiners differ from non-joiners within their counties, and the SS is distinct from the broader NSDAP.\footnote{We observe the near-universe of NSDAP membership, but the \textit{Ortsgruppenkartei} lost major holdings from Franconia (Section~\ref{sec:data_individual}). Our results are unaffected by dropping this region.}

\subsection{Fact 1: Discrete Waves of Entry}
Figure~\ref{fig:membership_time} shows NSDAP membership dynamics in our data. Until 1929, membership and its growth remained small. Following the onset of the Great Depression in 1929--1930, membership growth accelerated. The two largest surges in membership, in 1933 and 1937, were sharp and discontinuous rather than gradual. 

The 1933 wave coincided with the NSDAP's seizure of power, which marked a resolution of political uncertainty and a decisive shift in the expected costs and benefits of joining. The partial lifting of the \textit{Mitgliedersperre} in 1937 produced a second wave of mass entry.  The timing of both waves aligns with institutional and social changes in barriers to entry rather than with gradual ideological conversion or economic shocks. By the end of the war, cumulative membership had reached approximately 11 million.

Our membership records also document exits, consisting of voluntary departures, expulsions, and deaths (see Panel~A of Figure \ref{fig:membership_leavers_ss}). Some members did leave, particularly before regime consolidation, and expulsion rates were non-trivial. At its peak in mid-1931, there were around 3,800 monthly leavers. These exits should not, however, be interpreted as evidence of widespread public opposition to the regime. Compared to the stock of members, exit rates were small. Organized resistance was rare and exacted extreme personal costs. Most opposition took the form of what \citet{broszat1981hitler} termed \textit{Resistenz}, passive non-compliance, foot-dragging, and cultural dissent, rather than active resistance \citep{peukert1987inside, kershaw2000nazi}.

\subsection{Fact 2: Granular Geographic Variation}
\begin{figure}[!t]
\caption{Geographic Distribution of NSDAP Members}
\centering
\includegraphics[width=\textwidth]{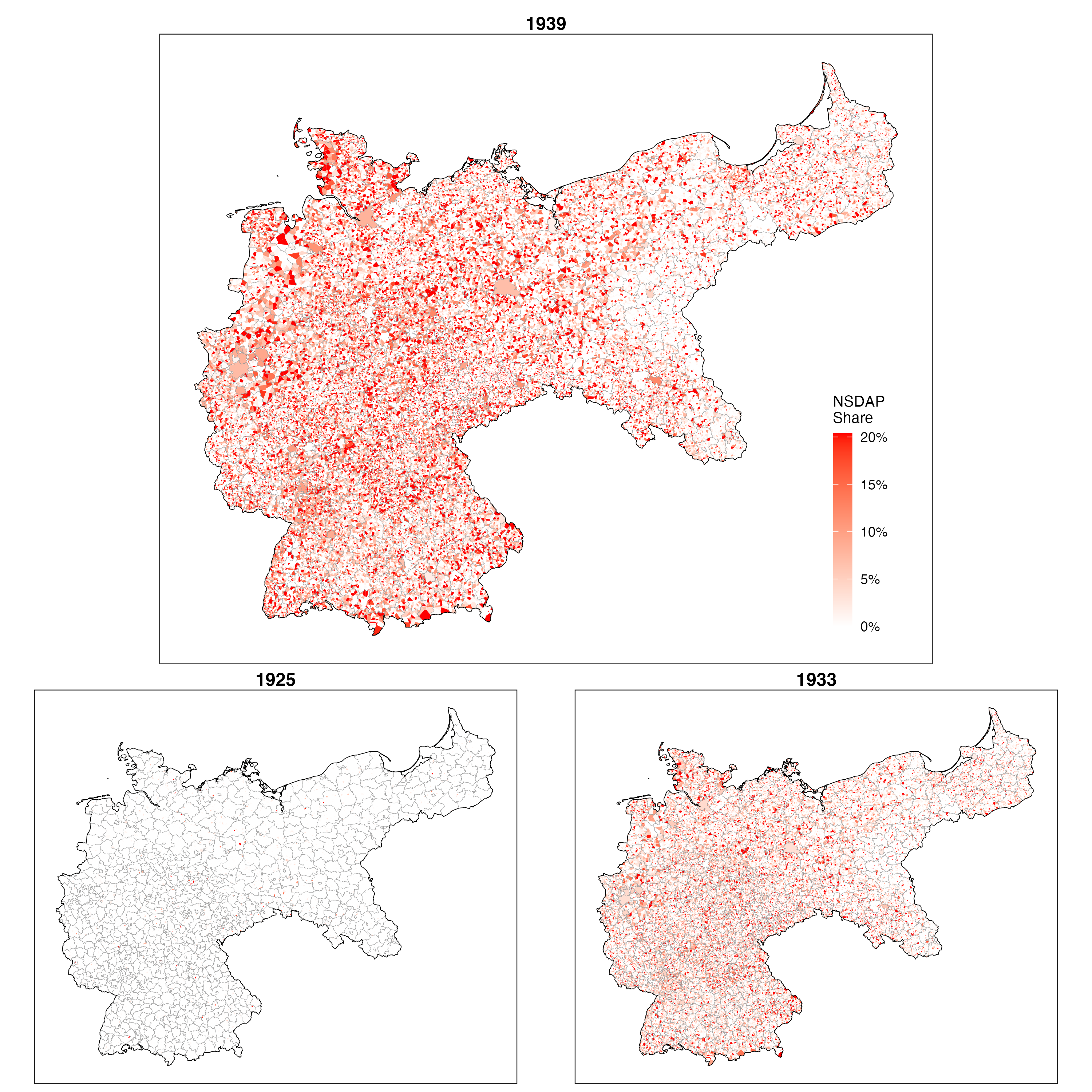}
\label{fig:map}
\\[4pt]
\justifying
\scriptsize{\textbf{Note.} This figure shows the share of NSDAP members in each municipality’s total population at three census dates. Municipal boundaries are constructed using Voronoi tessellation within county borders (see the Data Appendix). The color scale ranges from white (zero) to red (maximum share). County borders are overlaid in gray. Municipalities within the same county often exhibit substantial differences in membership shares, a pattern quantified in Figure~\ref{fig:variance_decomposition}.}
\end{figure}

Figure~\ref{fig:map} leverages the granularity of our data to show the geographic distribution of NSDAP members at the municipal level for three census years (1925, 1933, and 1939). Even within the same county, neighboring municipalities differ markedly in membership shares. The spatial pattern exhibits no clear gradient; areas with high and low membership shares coexist across the entire territory. Tracking this distribution over time reveals a high degree of persistence: municipalities that were early strongholds tend to remain so throughout the period (also see Appendix Figure~\ref{fig:membership_pre1933}). 

\begin{figure}[!t]
\caption{Within-County vs.\ Between-County Variation in NSDAP Membership}
\vspace{-.2cm}
\centering
\includegraphics[width=\textwidth]{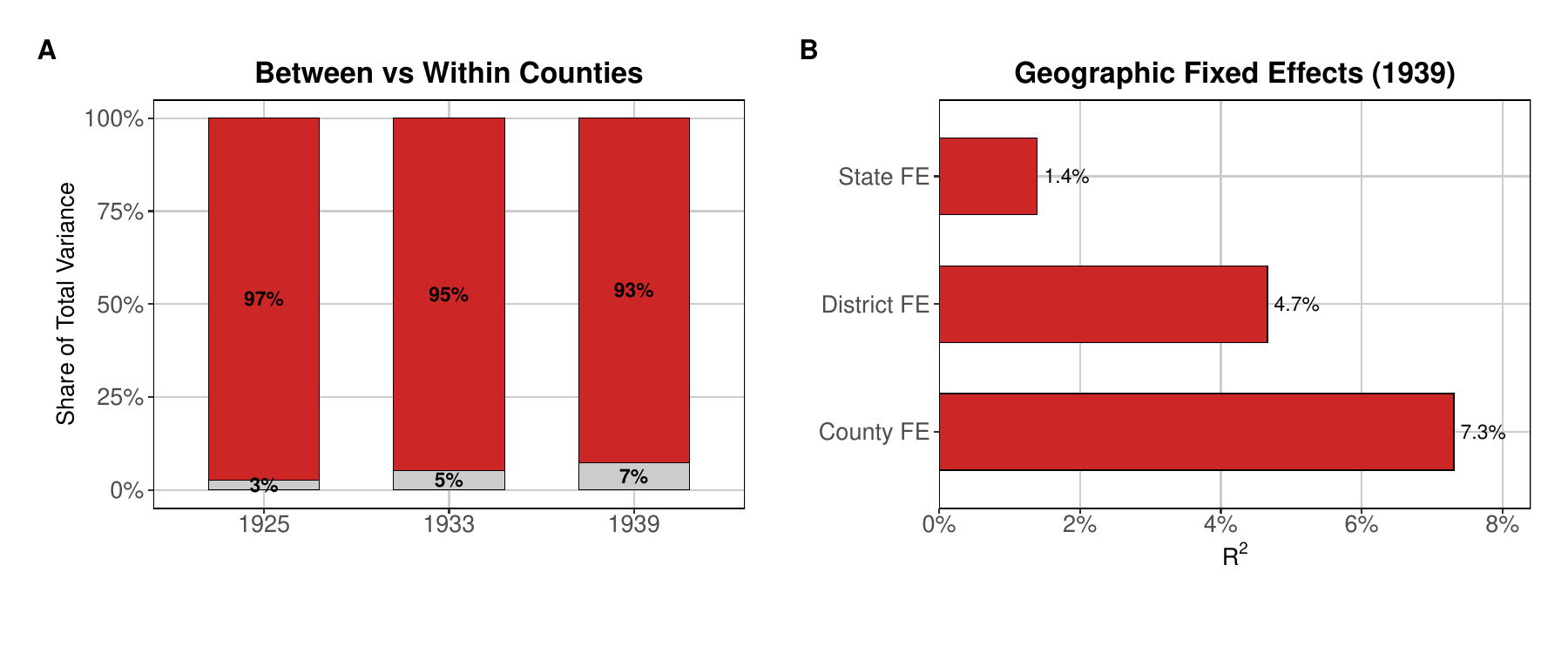}
\label{fig:variance_decomposition}
\\[4pt]
\vspace{-1.5cm}
\justifying
\scriptsize{\textbf{Note.} Panel~\textbf{A} decomposes the variance of municipal NSDAP membership shares into between-county (red) and within-county (gray) components at each of the three census dates. The reported statistic is the ratio of the between-county sum of squares to the total sum of squares. Panel~\textbf{B} reports the $R^2$ from regressions of municipal membership shares in 1939 on progressively finer sets of geographic fixed effects.}
\end{figure}

Figure~\ref{fig:variance_decomposition} quantifies the importance of this within-county variation. Panel~A decomposes the total variance in municipal NSDAP membership shares into a between-county and a within-county component at three census dates. Around 95\% of the total variation occurs \textit{within} counties, a share that is stable over time (97\% in 1925, 95\% in 1933, and 92\% in 1939). Panel~B is consistent with this finding and shows the results of cross-sectional regressions of municipal membership shares on progressively finer geographic fixed effects in 1939. State fixed effects explain less than 1\% of the variation; even county fixed effects, the finest administrative unit in the prior literature, explain only about 3\%.

This finding has a direct methodological implication. The existing literature on NSDAP support has relied almost exclusively on county-level data \citep{falter1991hitlers, king1997solution}, which by construction captures only the between-county component. Our results show that this approach misses the majority of the variation in party penetration. Individual-level data are necessary to study the mechanisms that generated heterogeneous mobilization within the same local institutional environment.

Local party penetration had consequences beyond politics: municipalities with higher NSDAP membership subsequently experienced higher rates of deportation and emigration during the Second World War, a relationship we examine in Section~\ref{sec:consequences}.

\subsection{Fact 3: Convergence to the Population}
We next examine whether the composition of joiners changed as the NSDAP grew from a fringe movement into a mass organization. For each county $c$, entry period $t$, and characteristic $k$, let $x_{ctk}^{g}$ denote the share of characteristic $k$ among entrants into group $g$ in period $t$, and let $x_{ck}$ denote the corresponding county population share from the relevant census benchmark. In this subsection, $g=N$ denotes the NSDAP. We measure the distance between entrants and the county population across the set of characteristics $\mathcal{K}$. We use the Euclidean distance

\begin{equation}
D_{ct}^{g} = \left( \frac{1}{|\mathcal{K}|} \sum_{k \in \mathcal{K}} \left(x_{ctk}^{g} - x_{ck}\right)^2 \right)^{1/2},
\label{eq:divergence}
\end{equation}
where higher values indicate that entrants differ more from the local population. We then average this distance across counties within each period,
\begin{equation}
\bar{D}_t^{g} = \frac{1}{|\mathcal{C}_t^{g}|} \sum_{c \in \mathcal{C}_t^{g}} D_{ct}^{g},
\end{equation}
where $\mathcal{C}_t^{g}$ is the set of counties with at least ten entrants into group $g$ in period $t$. Finally, we rescale the resulting series to the unit interval.

\begin{figure}[!t]
\caption{Representativeness of Joiners Over Time}
\vspace{-.2cm}
\centering
\includegraphics[width=\textwidth]{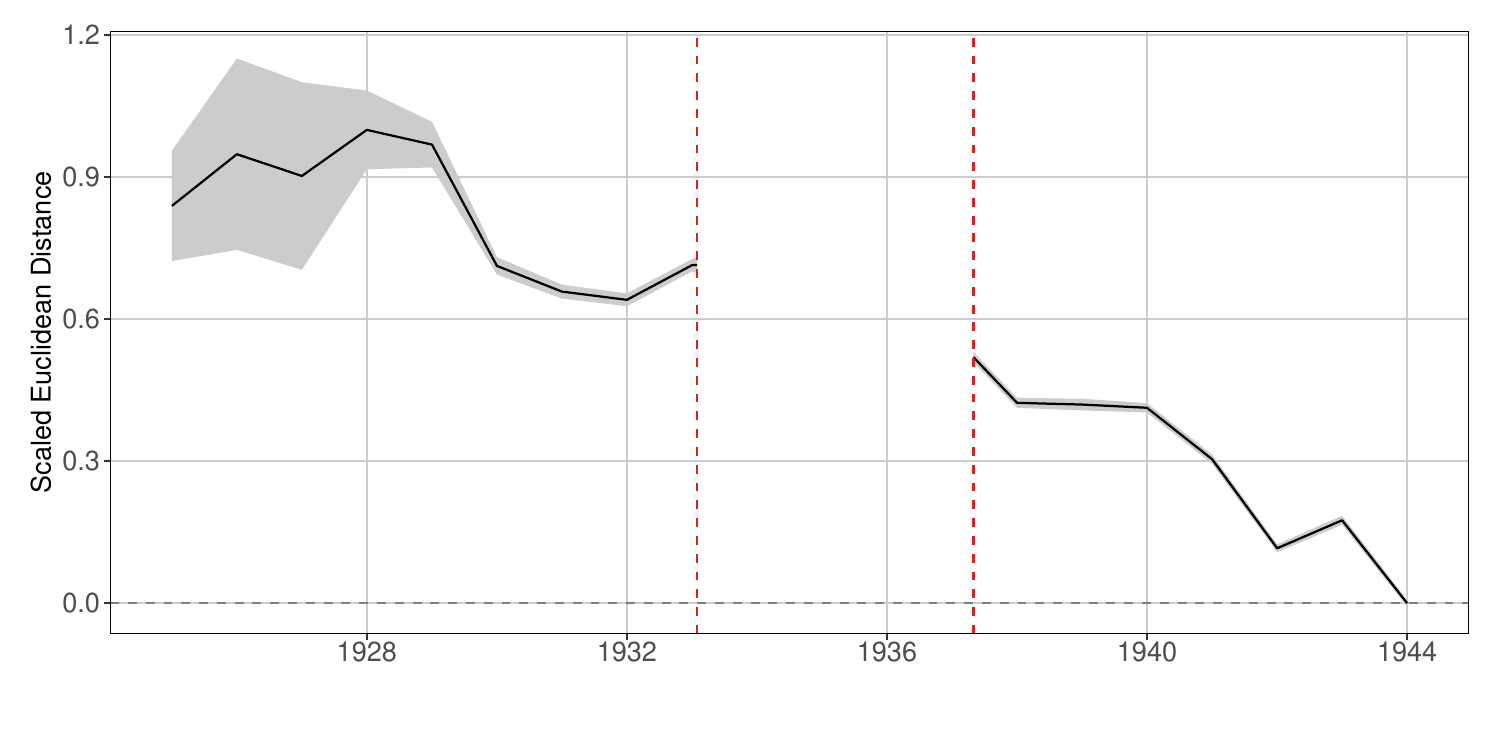}
\label{fig:county_time}
\\[4pt]
\vspace{-1.5cm}
\justifying
\scriptsize{\textbf{Note.} Average distance $\bar{D}_t$ between NSDAP joiners and their county population across eight characteristics (agricultural share, industrial share, trade share, administrative share, worker share, self-employed share, Protestant share, female share), averaged across counties with at least ten joiners per year and rescaled to $[0,1]$. Higher values indicate that joiners differ more from the local population. Vertical dashed lines mark January~30, 1933 and May~1, 1937 (admissions freeze). Shaded area shows 95\% confidence intervals.}
\end{figure}

Figure~\ref{fig:county_time} reports the results. In the party's early years (1925--1928), joiners differ substantially from their county populations. This divergence declines through the early 1930s, although there is a small uptick in 1933, when the NSDAP seized power. After the admissions freeze ended in 1937, divergence falls further, reaching its lowest level by the early 1940s. The pattern suggests that early joiners were a highly selected group. As the party expanded, new members increasingly resembled the broader local population, consistent with a transition from ideological vanguard to mass organization.

\subsection{Fact 4: Joiner Counties}

\begin{figure}[!t]
\caption{Community Composition: Between-County and Within-County Differences Before 1933}
\vspace{-.25cm}
\centering
\includegraphics[width=\textwidth]{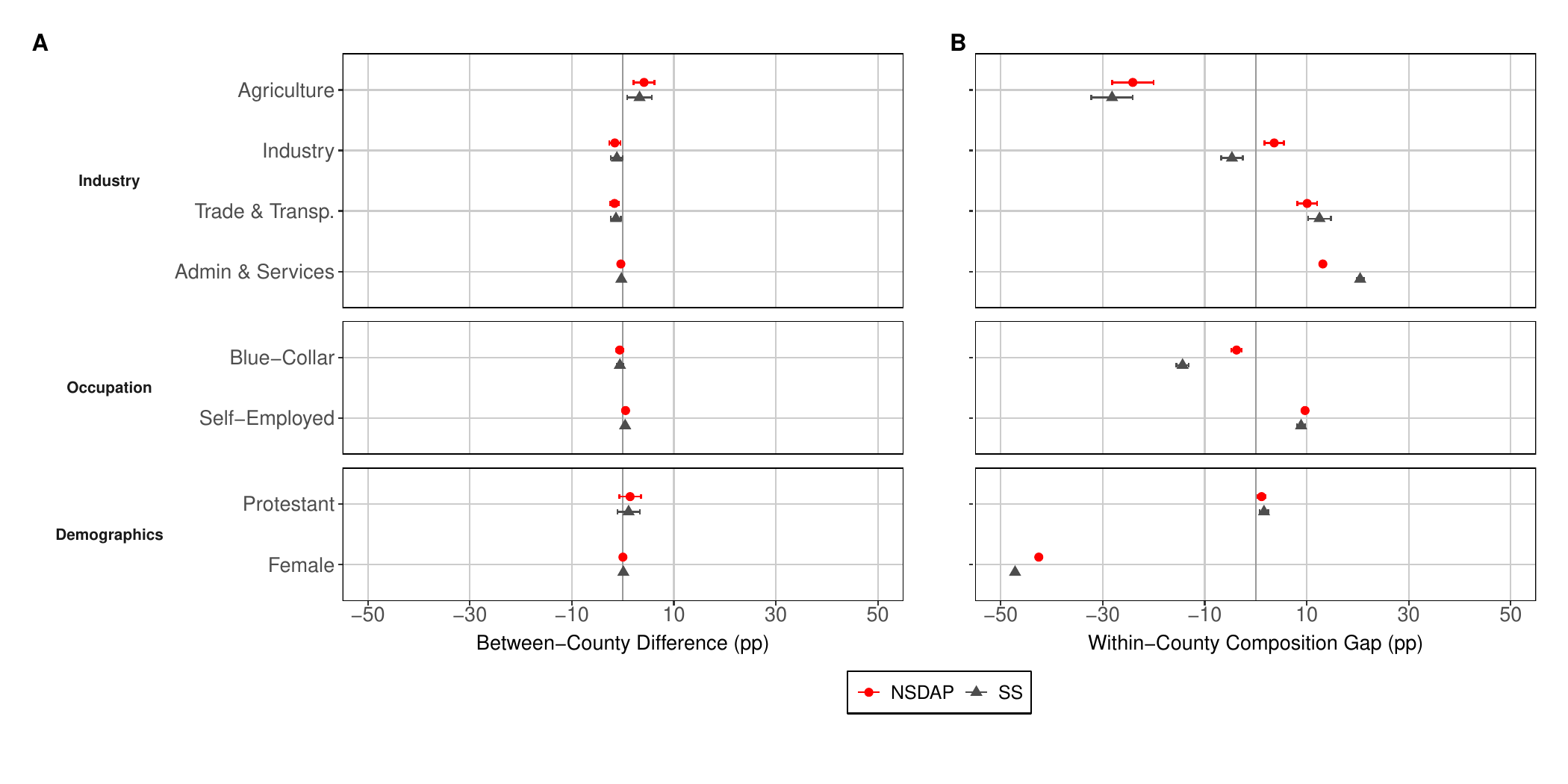}
\label{fig:community_butterfly}
\\[4pt]
\vspace{-1cm}
\justifying
\scriptsize{\textbf{Note.} Panel~\textbf{A} plots $\Delta_k^{g,\textup{between}}$: the group-weighted county average of characteristic $k$ minus the national aggregate share. The national aggregate is computed with denominator-consistent county weights $\omega_c^{\textup{pop}} = D_{ck}/\sum_{c'} D_{c'k}$, where $D_{ck}$ is the county workforce for occupational shares and total county population for demographic shares, so that $\sum_c \omega_c^{\textup{pop}} x_{ck}$ coincides with the national share $\sum_c N_{ck}/\sum_c D_{ck}$. Panel~\textbf{B} plots $\Delta_k^{g,\textup{within}}$: the group-weighted average gap between group composition ($x_{ck}^g$) and population composition ($x_{ck}$) within the same county. NSDAP ($g=N$) is shown in red and SS ($g=S$) in grey. All values are reported in percentage points. Error bars show 95\% confidence intervals with standard errors clustered at the county level.}
\end{figure}

How do the communities from which NSDAP and SS members were drawn differ from the broader German population? Figure~\ref{fig:community_butterfly} decomposes the total difference into a between-county and a within-county component. For each group $g \in \{N,S\}$ (NSDAP and SS), we examine the same set of characteristics $k \in \mathcal{K}$. We consider three sets of characteristics: sector of employment (agriculture, industry, trade and transport, administration and services), occupational status (blue-collar and self-employed), and demographic characteristics (Protestant and female).

Panel~A reports between-county differences. Let $x_{ck} = N_{ck}/D_{ck}$ denote the census share of characteristic $k$ in county $c$ on its natural base $D_{ck}$ (the county workforce for occupational shares; total county population for demographic shares). Let $M_c^g$ denote the number of group-$g$ members from county $c$. Define group and population county weights $\omega_c^g = M_c^g / \sum_{c'} M_{c'}^g$ and $\omega_c^{\textup{pop}} = D_{ck} / \sum_{c'} D_{c'k}$. The between-county component is
\begin{equation}
\Delta_k^{g,\textup{between}} = \sum_c \omega_c^g x_{ck} - \sum_c \omega_c^{\textup{pop}} x_{ck},
\label{eq:between}
\end{equation}
where, by construction, the second term is the aggregate national share of characteristic $k$ over the counties on common support $\sum_c \omega_c^{\textup{pop}} x_{ck} = \sum_c N_{ck}/\sum_c D_{ck}$, i.e.\ the value an average German on the relevant base would exhibit. The between-county component is therefore positive when group members are disproportionately drawn from counties in which characteristic $k$ is more prevalent than in the national aggregate. Appendix Figure~\ref{fig:community_butterfly_unweighted} reports results of an unweighted approach as a robustness exercise.

Panel~A of Figure \ref{fig:community_butterfly} shows that differences between NSDAP-weighted and population-weighted means are modest for most characteristics, consistent with \citeauthor{kater1983nazi}'s (\citeyear{kater1983nazi}) emphasis on the old \textit{Mittelstand} (small business owners, independent artisans, and farmers).\footnote{Appendix Figure~\ref{fig:falter_representativeness} applies our coding to the 0.5\% membership file compiled by \citet{falter1991hitlers} and \citet{kater1983nazi}. Our weighted marginal shares reproduce Kater's published aggregates within classification-judgement tolerance. At the cluster-design ceiling for their 0.5\% sampling fraction, however, the population-weighted between-county gap is not sign-identified for any of the eight characteristics, while the unweighted cross-county gap is sign-identified for all eight. The prior literature's reliance on the unweighted benchmark reflects this identification constraint of the 0.5\% sample rather than a modeling choice; the full-population estimates reported here flip the sign of Agriculture, Industry, and Trade~\&~Transport relative to that benchmark.} The same holds for SS members.\footnote{Appendix Figure~\ref{fig:community_butterfly_unweighted} shows the unweighted-benchmark analogue, in which NSDAP members appear disproportionately drawn from more Protestant and less agricultural counties. This pattern reflects the fact that small rural counties and the counties from which NSDAP members are drawn are both over-represented when counties are weighted equally, rather than a genuine population-share difference.}%

\begin{figure}[!t]
\caption{Geographic Differences Between Counties, 1933}
\vspace{-.25cm}
\centering
\includegraphics[width=.8\textwidth]{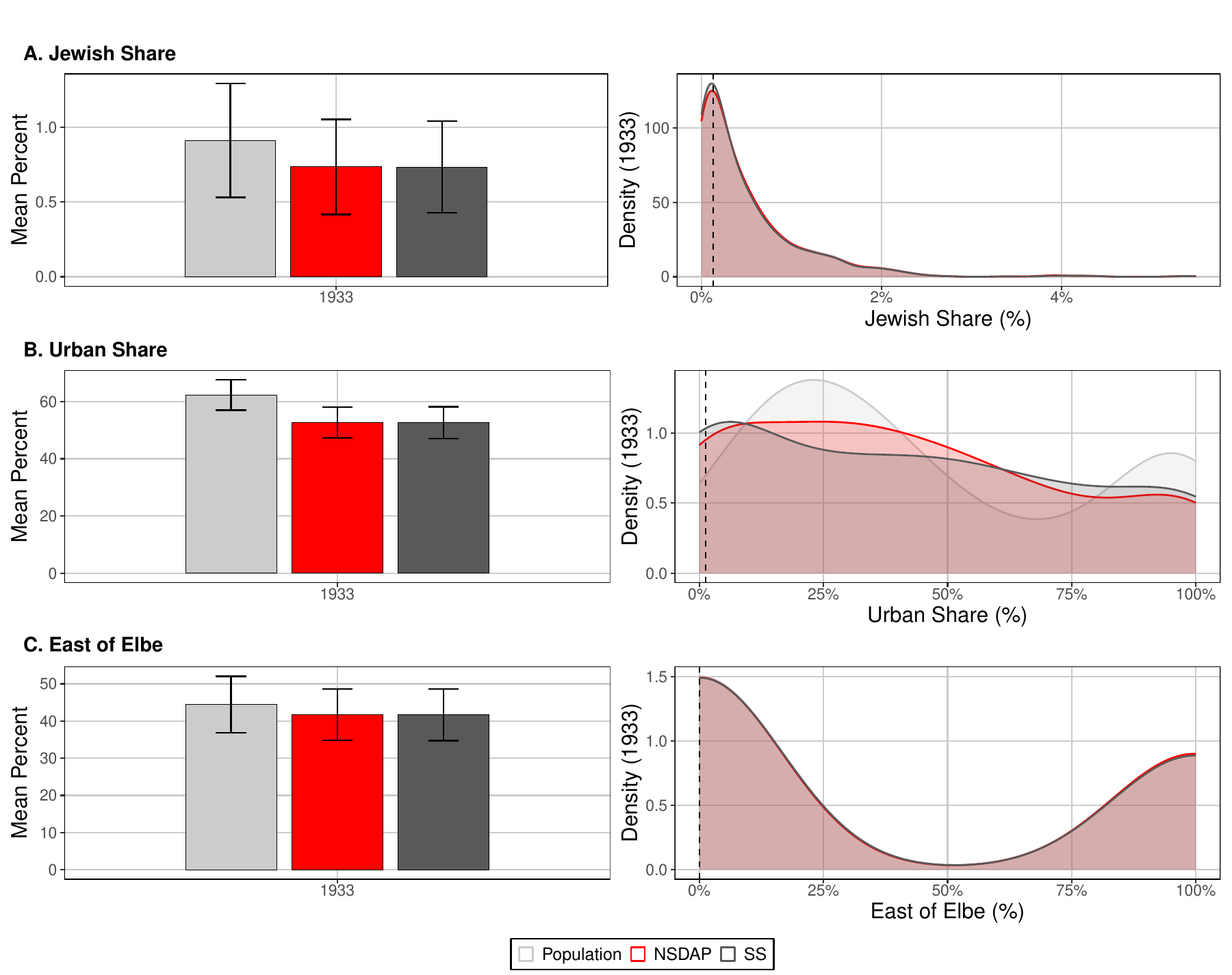}
\label{fig:dashboard_geographic}
\\[4pt]
\justifying
\scriptsize{\textbf{Note.} Each row reports one county-level geographic feature: Jewish population share (A), urban share (B), and East-of-Elbe location (C). Left panels show weighted county-level means in 1933 for the general population (grey), NSDAP members (red), and SS members (dark grey); error bars show 95\% confidence intervals with standard errors clustered at the county level. Right panels overlay county-level kernel densities for 1933; the dashed vertical line marks the location of the largest pairwise Kolmogorov--Smirnov gap among the three pairwise group comparisons.}
\end{figure}

Figure~\ref{fig:dashboard_geographic} extends the same three-group comparison to three additional features for different census years: county Jewish population share, urban status, and East-of-Elbe location.\footnote{We do not disaggregate Panel~A in Figure \ref{fig:community_butterfly} variable by variable, because the between-county component varies across census dates only through the evolution of group member counts and therefore does not lend itself to the three-year bar format used here.} The left column reports weighted county-level means in 1933 with 95\% confidence intervals clustered at the county level; the right column overlays the 1933 county-level densities and marks the largest pairwise Kolmogorov--Smirnov gap with a dashed line. NSDAP and SS members are drawn from counties that are marginally less Jewish, slightly less urban, and essentially on par with the national aggregate on East--West location. None of the differences are statistically significant. The 1933 distributions track the population benchmark closely across all three features.

\subsection{Fact 5: Joiner Characteristics}
We next examine whether joiners and non-joiners differ \textit{within} their communities. To this end, we compare county-level characteristics to the aggregate characteristics of NSDAP or SS members within the same county. We draw on the same dimensions as in Figure~\ref{fig:community_butterfly}, Panel~A, but now measured as individual characteristics from the membership cards, comparing to county aggregates.

Figure~\ref{fig:community_butterfly}, Panel~B reports within-county differences. Let $x_{ck}^g$ denote the share of characteristic $k$ among group-$g$ members in county $c$, computed from individual membership records. The within-county component is
\begin{equation}
\Delta_k^{g,\textup{within}} = \sum_c \omega_c^g \bigl(x_{ck}^g - x_{ck}\bigr).
\label{eq:within}
\end{equation}
This quantity is positive when, within the same county, group members are more likely to have characteristic $k$ than the general population. Standard errors are clustered at the county level. The two components sum to the total deviation of the group mean from the population average:
\[
\Delta_k^{g,\textup{between}} + \Delta_k^{g,\textup{within}} = \sum_c \omega_c^g x_{ck}^g - \sum_c \omega_c^{\textup{pop}} x_{ck}.
\]
Results indicate that within-county selection dominates between-county sorting. The within-county components (Figure~\ref{fig:community_butterfly}, Panel~B) substantially exceed the between-county components (Panel~A) for most characteristics. Within their counties, NSDAP and SS members are drawn disproportionately from non-agricultural occupations and are overrepresented in industry, trade, and especially administration and services. The self-employed are modestly overrepresented among NSDAP members, consistent with the historical emphasis on the \textit{Mittelstand}. The female-share gap is large, reflecting the overwhelmingly male composition of both organizations.\footnote{Note that while the NSDAP was overwhelmingly composed of men, the baseline census data include women. Because occupational structure varies by gender, comparing a male-dominated cohort to the raw general population can create a distorted metric for ordinariness. The census data report all female employment by industry and occupation, which allows us to account for this.} Within counties, there is also no confessional gap, while occupational and gender gaps are large. The SS exhibits a broadly similar pattern, but SS members are substantially more likely to work in public administration and even less likely to hold blue-collar occupations.

\begin{figure}[!t]
\caption{NS Symbols in Portraits: NSDAP vs.\ SS}
\centering
\includegraphics[width=\textwidth]{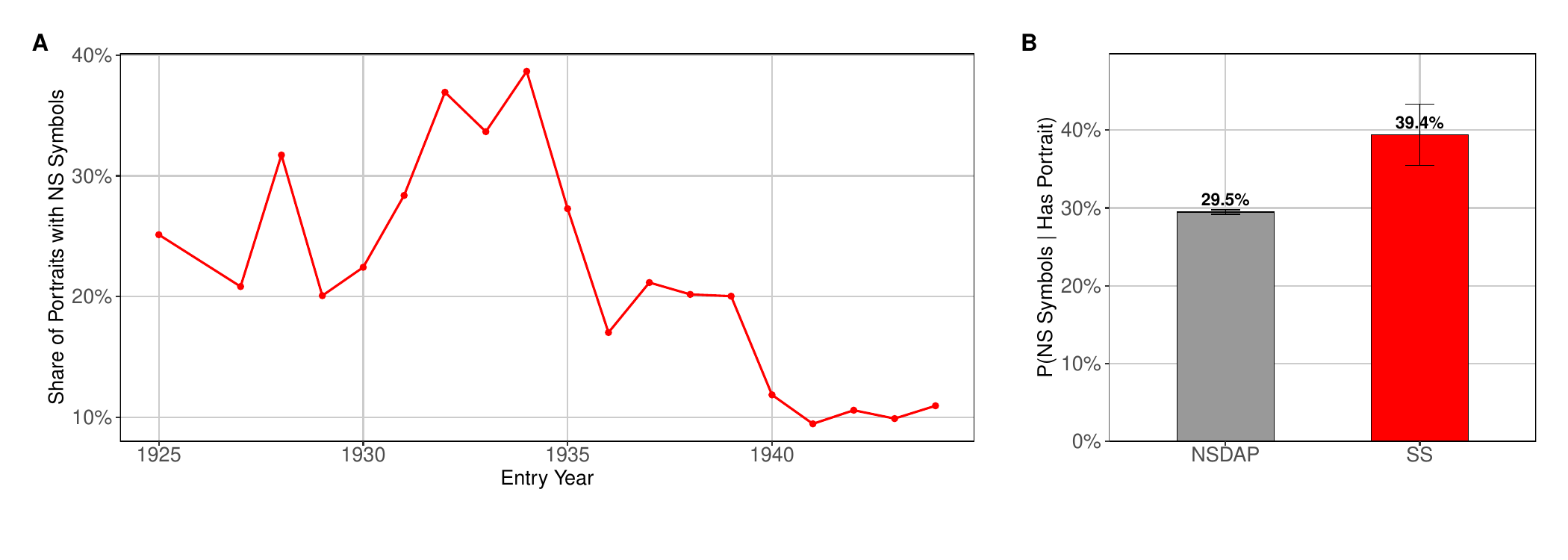}
\label{fig:ns_symbols_ss}
\\[4pt]
\vspace{-1cm}
\justifying
\scriptsize{\textbf{Note.} Panel~A shows the share of NSDAP membership portraits displaying NS symbols by individual entry year. Each portrait is counted in the year corresponding to its member's entry. Panel~B shows the probability of displaying NS symbols conditional on having a portrait, separately for NSDAP and SS members. Sample sizes shown on bars.}
\end{figure}

\subsection{Fact 6: The SS Selection}

While both NSDAP and SS members differed in some respects from the broader population, this comparison masks large differences between the two groups that are not captured in the census aggregates. SS members were drawn disproportionately from a specific demographic: young men born between approximately 1905 and 1912, with above-average education, often holding university degrees or doctorates, and employed in academic, legal, or medical professions (see Appendix Figure \ref{fig:nsdap_ss}). 

We also try to capture potential differences in ideological alignment with the Nazi regime. To do so, we analyze all available membership portraits and record whether members wear NS insignia in their photographs. Panel A of Figure~\ref{fig:ns_symbols_ss} shows that, until 1939, between 20 and nearly 40\% of new NSDAP members displayed some NS insignia in their membership file portrait. Panel B shows that SS members were about 10 percentage points more likely to do so than other NSDAP members.

These empirical patterns are consistent with historical accounts suggesting that the SS functioned as a self-conscious elite, with Himmler cultivating a ``New Aristocracy from Blood and Soil'' through selective recruitment and rigorous screening \citep[p. 127]{longerich2012heinrich}. 

\section{Coordination Dynamics}
\label{sec:coordination}

Section~\ref{sec:descriptives} documented two patterns: entry into the NSDAP occurred in discrete waves, and most variation in membership was within counties rather than between them. We now examine to what extent coordination dynamics within communities can account for these patterns.

\begin{figure}[!t]
\caption{Share of Municipalities with NSDAP Members}
\centering
\includegraphics[width=1.0\textwidth]{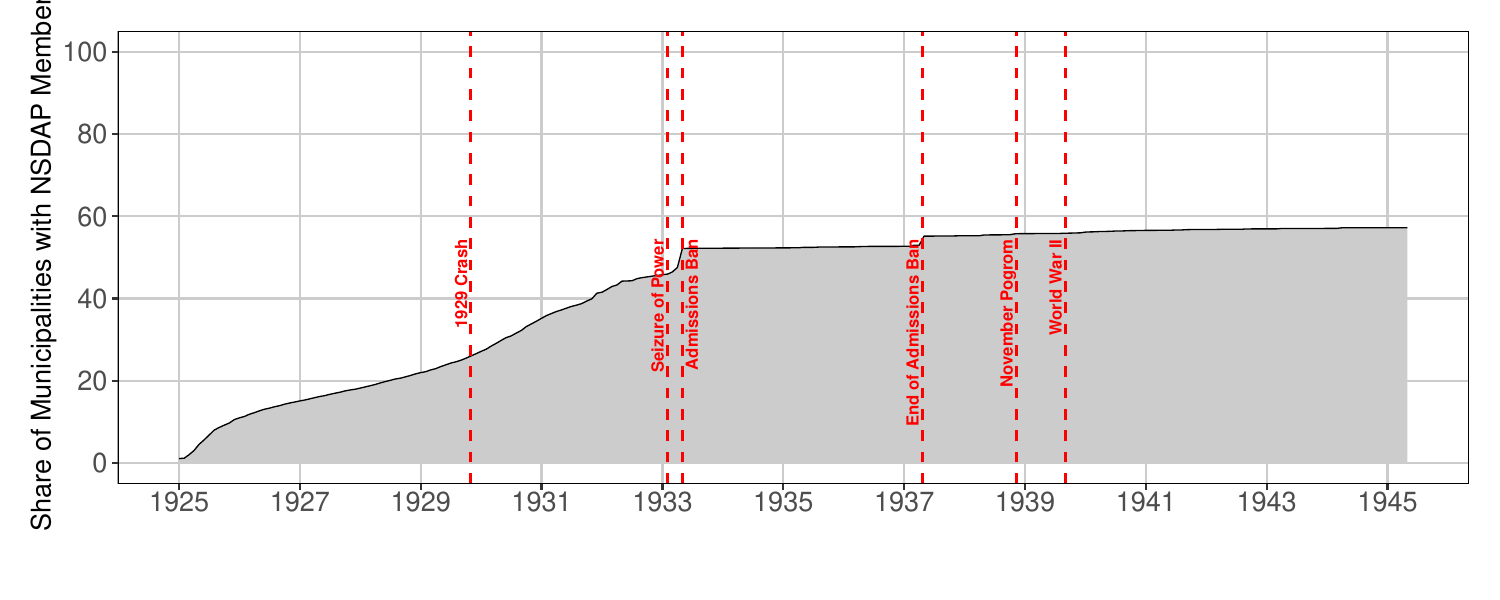}
\label{fig:coordination_zero_share}
\\[4pt]
\vspace{-1cm}
\justifying
\scriptsize{\textbf{Note.} Monthly share of municipalities with at least one cumulative NSDAP member from 1925 to 1945. Each data point is the fraction of all municipalities in which at least one membership card records an entry date on or before that month.}
\end{figure}

Figure~\ref{fig:coordination_zero_share} plots the share of municipalities with at least one NSDAP member over time. Around half of all municipalities had no recorded members even in January 1933. Despite the sharp surge in membership after March 1933, the share of municipalities with any members rises only modestly from 48 to 56 percent by 1939. The mobilization wave was thus concentrated on the intensive margin: new members entered predominantly in municipalities where members were already present.

\begin{figure}[!t]
\caption{Coordination Dynamics in Municipal Entry}
\vspace{-.3cm}
\centering
\includegraphics[width=0.85\textwidth]{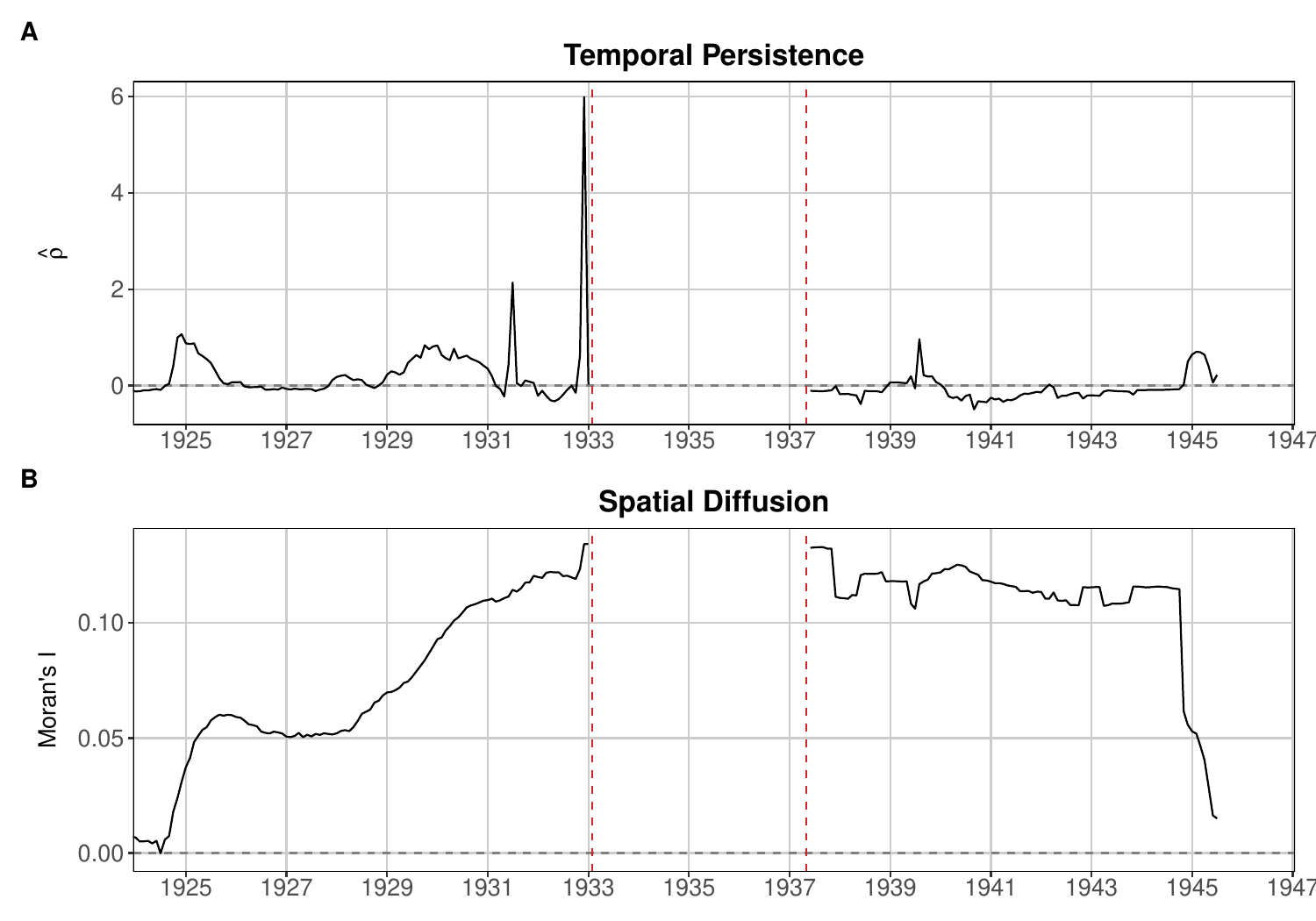}
\label{fig:coordination_dynamics}
\\[4pt]
\justifying
\vspace{-.3cm}
\scriptsize{\textbf{Note.} Panel~A reports rolling 12-month estimates of the AR(1) coefficient $\hat{\rho}$ from monthly municipal entry regressions with municipality and month fixed effects. Panel~B reports rolling 12-month Moran's~$I$ for log municipal entry using 10-nearest-neighbor weights across municipalities. Vertical dashed lines mark the admissions ban from May~1, 1933 to May~1, 1937.}
\end{figure}

We test this intensive-margin pattern more formally. Panel~A of Figure~\ref{fig:coordination_dynamics} plots rolling 12-month estimates of $\rho$ from
\[
n_{mt} = \alpha_m + \gamma_t + \rho\, n_{m,t-1} + \varepsilon_{mt},
\]
where $n_{mt}$ is monthly entry in municipality~$m$. The municipality fixed effects $\alpha_m$ absorb permanent cross-municipality differences, and $\hat{\rho}$ captures within-municipality temporal persistence. The coefficient is near zero through the 1920s, spikes sharply around 1933, and then falls back. The 1933 spike indicates strong serial correlation in within-municipality entry. 

Next, we turn to spatial auto-correlation. Panel~B reports Moran's~$I$ for log monthly entry using nearest-neighbor weights. Moran's~$I$ captures spatial dependence between municipalities, and we find that it rises gradually through the late 1920s and early 1930s and remains elevated after 1937, but always remains at modest levels.

This contrast reinforces the intensive-margin interpretation: within-municipality temporal persistence is far stronger than cross-municipality spatial diffusion. $\hat{\rho}$ rises sharply around 1933, reflecting explosive within-municipality dynamics in which a single entry is followed by many additional entries in the same municipality in subsequent months. Moran's~$I$, by contrast, remains modest throughout and builds gradually rather than in discrete waves. 

Persistence also extends across the admissions ban: municipalities that mobilized before May~1933 mobilized again at near-proportional rates after May~1937 (Appendix Figure~\ref{fig:cross_wave_persistence}).

We next ask what happens within a municipality after its first joiner. We define three network cells: \textit{community} (the municipality); \textit{workplace} (the municipality-by-industry-by-occupational-status cell); and \textit{family} (individuals sharing a surname within the same municipality).\footnote{These cells are defined ecologically rather than by documented interpersonal ties: shared surnames within a municipality need not indicate kinship, and workers in the same municipality-by-industry-by-occupation-status cell need not share a physical workplace. Our estimates therefore may reflect more than interpersonal networks alone, and we interpret them accordingly. At the same time, the median municipality has about 240 inhabitants; thus, in most municipalities, we believe this measure carries a meaningful signal of interpersonal network exposure.} For each cell~$c$, let $T_c$ denote the year of the first joiner and estimate
\begin{equation}
y_{ct} = \alpha_c + \gamma_t + \sum_{k=0}^{K}\beta_k\,\mathbf{1}\{t - T_c = k\} + \varepsilon_{ct},
\end{equation}
where $y_{ct}$ is the share of eventual cell joiners who enter in year~$t$. The denominator excludes the first joiner in year~$T_c$, so the coefficients trace the subsequent wave net of the triggering event. Because the outcome is mechanically zero before the first joiner, pre-event periods are shown empty and period~$-1$ serves as the reference. For the workplace and family channels, we restrict to cells with at least five eventual joiners. Standard errors are clustered at the county level.

\begin{figure}[!t]
\caption{Cascade Dynamics}
\vspace*{-0pt}
\centering
\includegraphics[width=.95\textwidth]{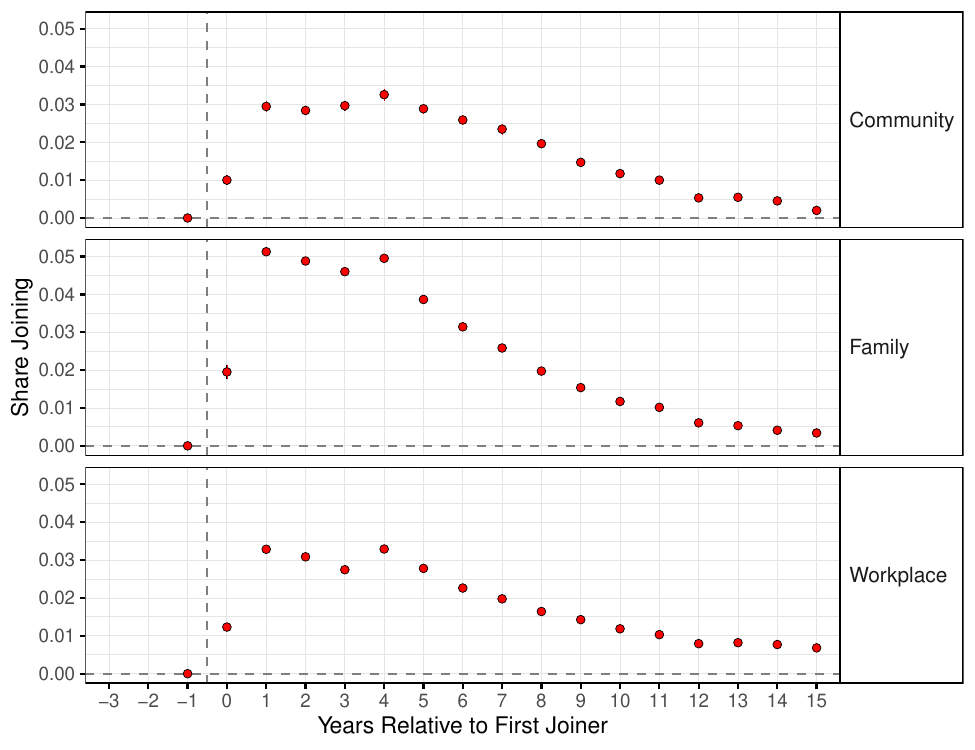}
\label{fig:community_cascade}
\\[4pt]
\justifying
\scriptsize{\textbf{Note.} Descriptive event studies at the network-cell level. The outcome is the share of eventual cell joiners who enter in year~$t$, with the first joiner removed from the denominator. \textit{Community}: municipality. \textit{Workplace}: municipality-by-industry-by-occupational-status cell, restricted to cells with at least five eventual joiners. \textit{Family}: individuals sharing a surname within the same municipality, restricted to cells with at least five eventual joiners. Event time is defined relative to the first cell joiner. Pre-event periods are mechanically zero and shown empty; period~$-1$ is the reference. Post-period coefficients are estimated with cell and year fixed effects. Standard errors are clustered at the county level.}
\end{figure}

Figure~\ref{fig:community_cascade} reports the results. In all three types of networks, the share of eventual joiners entering in a given post-event year is positive immediately after the first cell joiner and peaks within the first five years. Entry rates decline gradually afterward as the pool of remaining joiners shrinks. Once entry begins in a cell, further entry arrives in a sustained wave.

This evidence is descriptive. It does not identify a peer effect in the broader population, nor does it distinguish propagation from saturation or common local shocks. To explore the impact of saturation in more detail, we trace the share of eventual joiners already in the party against subsequent entry growth in a flow map in Appendix Figure~\ref{fig:flow_map_appendix}; growth is steep at first and flattens, consistent with saturation in the pool of eventual joiners. Appendix Figure~\ref{fig:early_later} is also consistent with this pattern: municipalities with higher cumulative membership through 1932 exhibit higher joining rates in 1933--1939. Together, the temporal persistence in Panel~A of Figure~\ref{fig:coordination_dynamics} and the post-entry wave in Figure~\ref{fig:community_cascade} indicate that local entry dynamics were self-reinforcing and highly localized within municipalities. These findings are consistent with recent quantitative case studies examining local entry dynamics \citep{caesmann2025viral}.

A central fact from our analysis is that entry into the NSDAP accelerated sharply in March~1933. We close the section by asking how this acceleration varied with pre-determined municipal characteristics. We construct a municipality-by-month panel covering January~1932 through June~1934 and estimate
\begin{equation}
\text{NSDAP share}_{mt} = \alpha_m + \gamma_t + \beta\,(\text{Post}_t \times Z_m) + \delta\,(\text{Post}_t \times \text{Any}_m) + \varepsilon_{mt},
\end{equation}
where $\text{Post}_t$ turns on in March~1933, $Z_m$ is a municipal characteristic computed from members who joined before March~1933, and $\text{Any}_m$ indicates the presence of any pre-March NSDAP member. We include $\text{Post}_t \times \text{Any}_m$ to estimate the coefficient net of pure party presence. Municipality and month fixed effects absorb local levels and the national trajectory; standard errors cluster by county. A causal interpretation of $\beta$ would require $Z_m$ to be uncorrelated with other drivers of post-March growth conditional on pre-March presence; this assumption is strong and we interpret $\beta$ as a descriptive association. We consider three cuts of $Z_m$: local notables (Table~\ref{table:gleich_notables}), demographic and industrial composition (Appendix Table~\ref{table:gleich_composition}), and paramilitary presence (Appendix Table~\ref{table:gleich_paramilitary}). The final column of the notables and composition tables enters all traits jointly.

Three patterns emerge. First, the post-March 1933 surge in entry is largest in municipalities whose pre-March base was disproportionately female (Appendix Table~\ref{table:gleich_composition}). Compositionally, the wave accelerated where the pre-existing NSDAP base was more agricultural and tilted toward civic-administrative occupations, not toward industry or the working class. Second, among local notables, we find the strongest differences for teacher, a group disproportionately targeted by Nazi propaganda \citep{kater1983nazi, falter2020parteigenossen, voth2015indoctrination} and central to the dissemination of local norms. Historically, education played an important role in fostering compliance \citep{paglayan2022education} and shaping ideological preferences \citep{alesina2021nation, cantoni2019curricula}. Beyond teachers, doctors and policemen retain a positive partial association in the joint specification. Pastors, industrialists, and estate owners are not distinguishable from zero in the bivariate specification and turn negative when entered jointly. Third, a pre-March 1933 SS and SA presence is associated with a substantial additional post-March increase over and above bare party presence (see Appendix Table~\ref{table:gleich_paramilitary}). This is consistent with an active coordination role for militant cadres and also with unobserved local characteristics generating both early SS recruits and a larger mass entry.

\begin{table}[!t]
\centering
\caption{The March 1933 Wave: Pre-March Local Notables}
\label{table:gleich_notables}
\resizebox{\textwidth}{!}{%

\begingroup
\centering
\begin{tabular}{lccccccc}
   \toprule
    & Teacher & Doctor & Policem. & Pastor & Indust. & Estate & Joint \\ \cmidrule(lr){2-2} \cmidrule(lr){3-3} \cmidrule(lr){4-4} \cmidrule(lr){5-5} \cmidrule(lr){6-6} \cmidrule(lr){7-7} \cmidrule(lr){8-8}
    & \multicolumn{7}{c}{NSDAP membership share}\\
                                      & (1)           & (2)           & (3)           & (4)           & (5)           & (6)           & (7)\\  
   \midrule 
   Post March 1933 x Any NSDAP Member & 0.03$^{***}$  & 0.03$^{***}$  & 0.03$^{***}$  & 0.03$^{***}$  & 0.03$^{***}$  & 0.03$^{***}$  & 0.03$^{***}$\\   
                                      & (0.0003)      & (0.0003)      & (0.0003)      & (0.0003)      & (0.0003)      & (0.0003)      & (0.0003)\\   
   Post March 1933 x Teacher          & 0.009$^{***}$ &               &               &               &               &               & 0.009$^{***}$\\   
                                      & (0.0006)      &               &               &               &               &               & (0.0007)\\   
   Post March 1933 x Doctor           &               & 0.006$^{***}$ &               &               &               &               & 0.003$^{***}$\\   
                                      &               & (0.0008)      &               &               &               &               & (0.0009)\\   
   Post March 1933 x Policeman        &               &               & 0.005$^{***}$ &               &               &               & 0.002\\   
                                      &               &               & (0.001)       &               &               &               & (0.001)\\   
   Post March 1933 x Pastor           &               &               &               & -0.001        &               &               & -0.006$^{***}$\\   
                                      &               &               &               & (0.002)       &               &               & (0.002)\\   
   Post March 1933 x Industrialist    &               &               &               &               & -0.0003       &               & -0.004$^{**}$\\   
                                      &               &               &               &               & (0.002)       &               & (0.002)\\   
   Post March 1933 x Estate owner     &               &               &               &               &               & -0.001        & -0.006$^{**}$\\   
                                      &               &               &               &               &               & (0.003)       & (0.003)\\   
   Municipality FE                    & \checkmark    & \checkmark    & \checkmark    & \checkmark    & \checkmark    & \checkmark    & \checkmark\\   
   Month FE                           & \checkmark    & \checkmark    & \checkmark    & \checkmark    & \checkmark    & \checkmark    & \checkmark\\   
   Mean Outcome                       & 0.0237        & 0.0237        & 0.0237        & 0.0237        & 0.0237        & 0.0237        & 0.0237\\  
    \\
   Observations                       & 1,761,720     & 1,761,720     & 1,761,720     & 1,761,720     & 1,761,720     & 1,761,720     & 1,761,720\\  
   R$^2$                              & 0.93327       & 0.93288       & 0.93281       & 0.93277       & 0.93277       & 0.93277       & 0.93334\\  
    \\
   \bottomrule
\end{tabular}
\par\endgroup

}
\\[4pt]
\justifying
\scriptsize{\textbf{Note.} Municipality-by-month panel, January~1932--June~1934. Dependent variable: municipal NSDAP membership share. Each column interacts $\text{Post March 1933}_t$ with a binary indicator equal to one if municipality~$m$ had at least one member of the indicated profession who joined before March~1933, controlling for $\text{Post}\times\text{Any NSDAP Member}$. Professions are classified from standardized occupation strings in the membership records; Doctor includes members with a doctorate or an \textit{Arzt} title excluding veterinarians and dentists. Column~7 enters all six notable interactions jointly. Municipality and month fixed effects included. Standard errors clustered at the county level in parentheses. Significance: $^{*}$~$p<0.10$, $^{**}$~$p<0.05$, $^{***}$~$p<0.01$.}
\end{table}

\section{Consequences}
\label{sec:consequences}

We next examine the consequences of local NSDAP penetration using two downstream outcomes: deportations and emigration. We construct a municipality-by-year panel covering all municipalities from 1930 to 1945. The dependent variables are annual counts of deportees and emigrants, drawn from the individual-level event records from the List of Jewish Residents in Germany 1933-1945, 1939 German Minority Census and the German Federal Archives' memorial book. We assign each victim to a municipality using their last recorded German residence. The regressor of interest is the NSDAP membership stock: the number of members whose entry date falls on or before the end of the year. We fix the Jewish population share at its 1933 level, since subsequent declines partly reflect the deportations we seek to explain. Standard errors are clustered at the county level. The estimating equation is
\begin{equation}
\operatorname{asinh}(Y_{mt}) \;=\; \alpha_m + \gamma_t + \beta \;\operatorname{asinh}\!\left(\frac{\text{NSDAP}_{mt}}{\text{pop}_m}\right) + X_{mt}'\psi + \varepsilon_{mt},
\end{equation}
where $Y_{mt}$ is the deportation or emigration count of Jews in municipality~$m$ and year~$t$, $\text{NSDAP}_{mt}$ is the stock of active members as of year-end, $\alpha_m$ and $\gamma_t$ are municipality and year fixed effects, and $X_{mt}$ includes log population, the 1933 Jewish population share, and an urban indicator. We use an inverse hyperbolic sine transformation for $Y_{mt}$ and $\text{NSDAP}_{mt}$. 

Table~\ref{table:consequences} reports the results. Columns~(1)--(3) examine deportations, columns~(4)--(6) emigration. Column~(1) and (4) include year fixed effects only. Column~(2) and (5) add county fixed effects, log population, the 1933 Jewish population share, and an urban indicator. Column~(3) and (6) replace these with municipality fixed effects and county-by-year fixed effects, so that identification comes entirely from within-municipality variation in NSDAP membership, comparing municipalities in the same county and year.

\begin{table}[hbt!]
\center
\begin{small}
\caption{Consequences of NSDAP Strength}
\begin{threeparttable}

\begingroup
\centering
\begin{tabular}{lcccccc}
   \toprule
    & \multicolumn{3}{c}{Deportations} & \multicolumn{3}{c}{Emigration} \\ \cmidrule(lr){2-4} \cmidrule(lr){5-7}
                              & (1)           & (2)           & (3)           & (4)           & (5)           & (6)\\  
   \midrule 
   NSDAP Members              & 0.006$^{***}$ & 0.006$^{***}$ & 0.009$^{***}$ & 0.003$^{**}$  & 0.003$^{***}$ & 0.003$^{**}$\\   
                              & (0.001)       & (0.001)       & (0.002)       & (0.001)       & (0.001)       & (0.001)\\   
   Controls                   &               & \checkmark    & \checkmark    &               & \checkmark    & \checkmark\\   
   Year FE                    & \checkmark    & \checkmark    & \checkmark    & \checkmark    & \checkmark    & \checkmark\\   
   County FE                  &               & \checkmark    &               &               & \checkmark    & \\  
   Municipality FE            &               &               & \checkmark    &               &               & \checkmark\\   
   Mean Outcome               & 0.005         & 0.005         & 0.005         & 0.003         & 0.003         & 0.003\\  
    \\
   Observations               & 1,386,399     & 1,386,399     & 1,386,399     & 1,386,399     & 1,386,399     & 1,386,399\\  
   R$^2$                      & 0.01048       & 0.09842       & 0.15746       & 0.00281       & 0.24714       & 0.31319\\  
    \\
   \bottomrule
\end{tabular}
\par\endgroup

\begin{tablenotes}
\item \scriptsize{\textbf{Note.} Municipal panel, 1930--1945. Dependent variable: $\operatorname{asinh}(\text{deportation count})$ in columns~(1)--(3), $\operatorname{asinh}(\text{emigration count})$ in columns~(4)--(6). Regressor: $\operatorname{asinh}(\text{NSDAP membership share})$. Controls where indicated: log population, Jewish population share (1933 baseline), urban indicator. Standard errors clustered at the county level. Significance: $^{*}$~$p<0.10$, $^{**}$~$p<0.05$, $^{***}$~$p<0.01$.}
\end{tablenotes}
\end{threeparttable}
\label{table:consequences}
\end{small}
\end{table}

Across all specifications, municipalities with higher NSDAP membership shares exhibit higher levels of deportation in the same year. The estimated association is positive and statistically significant in every specification, including the model with municipality and county-by-year fixed effects. NSDAP membership is also positively associated with Jewish emigration, consistent with the idea that stronger local party presence contributed to an environment in which Jewish residents were more likely to leave before deportations began.

We interpret these results as suggestive evidence that our membership data capture meaningful variation in local party presence that is predictive of regime violence. However, unobserved local ideology or institutional capacity could drive both NSDAP membership and persecution. The identifying assumption underlying column~(3) and~(6) is that, conditional on municipality and county-by-year fixed effects, changes in NSDAP membership are uncorrelated with other determinants of deportations. This assumption would be violated if, for example, local antisemitic sentiment simultaneously drove party growth and deportation intensity within the same municipality over time.

\section{Conclusion}
\label{sec:conclusion}

We digitize and analyze the near-universe of NSDAP membership records to study the rise of a mass extremist movement. As the party grew, members increasingly resembled the general population, with the composition gap narrowing after each wave of mass entry. Entry at the municipality level occurred in self-reinforcing waves: within-municipality temporal persistence in monthly entry rose around 1933 and again after the 1937 lifting of the admissions freeze. 

SS members remained notably distinct from the population. This contrast suggests the Nazi movement operated through two distinct channels. A broad coordination dynamic drew millions of ordinary Germans into the party. A narrower ideological selection mechanism channeled younger, more educated, and visibly more committed men into the regime's most extreme institutions. 

More broadly, our findings suggest that mass radicalization need not require a population composed mainly of ideological extremists. Under conditions of political uncertainty, entry into an extremist movement can spread through ordinary social settings and become self-reinforcing. In that sense, our interpretation is consistent with \citeauthor{browning1992ordinary}'s (\citeyear{browning1992ordinary}) account of Reserve Police Battalion~101 and with \citeauthor{arendt1951origins}'s (\citeyear{arendt1951origins}) argument that totalitarian movements draw strength from the masses as well as from a committed core.

Several caveats apply. The network estimates capture shifts in entry timing among eventual members and are built from observable proxies for social exposure, not direct measures of social ties. The family and workplace channels in particular rely on proxy cells with imperfect denominator information in the membership-only data. And while we document that municipalities with more NSDAP members subsequently experienced more deportations and Jewish emigration, the mechanism linking membership to regime violence remains to be identified.

These records, among the richest individual-level archives of any political movement in history, open the study of the rise of fascism to quantitative analysis at the level of individual decisions. They also open several avenues for future work, including denazification, the long-run career effects of party membership, and the mechanisms through which the SS recruited from the broader party base.

\newpage

\singlespacing

\printbibliography

\clearpage


\begin{refsection}

\appendix
\clearpage
\setcounter{table}{0}
\setcounter{figure}{0}
\setcounter{footnote}{0}
\setcounter{section}{0}

\setcounter{page}{1}
\renewcommand\thepage{A.\arabic{page}}

\renewcommand{\thesection}{\Alph{section}}
\renewcommand\thetable{\Alph{section}.\arabic{table}}
\renewcommand\thefigure{\Alph{section}.\arabic{figure}}

\section{Data}
\label{app:data}

\begin{table}[hbt!]
\center
\begin{small}
\caption{Summary Statistics}
\begin{threeparttable}
\resizebox{.8\textwidth}{!}{%
\begin{tabular}{lrrrrrr}
\hline\hline
 & N & Mean & SD & P25 & Median & P75 \\
\hline
\multicolumn{7}{l}{\textit{Panel A: Individual}} \\
\hline
Female & 10,944,717 & 0.145 & 0.352 & 0.000 & 0.000 & 0.000 \\
Protestant Prob. & 10,944,717 & 0.645 & 0.350 & 0.371 & 0.783 & 0.948 \\
Urban & 10,944,717 & 0.306 & 0.461 & 0.000 & 0.000 & 1.000 \\
East of Elbe & 10,944,717 & 0.340 & 0.474 & 0.000 & 0.000 & 1.000 \\
PhD & 10,944,717 & 0.009 & 0.097 & 0.000 & 0.000 & 0.000 \\
Nobility & 10,944,717 & 0.007 & 0.082 & 0.000 & 0.000 & 0.000 \\
SS Member & 39,654 & 1.000 & 0.000 & 1.000 & 1.000 & 1.000 \\
Age at Entry & 10,124,862 & 33.653 & 12.658 & 23.000 & 32.000 & 43.000 \\
Blue Collar & 8,777,905 & 0.388 & 0.487 & 0.000 & 0.000 & 1.000 \\
White Collar & 8,777,905 & 0.362 & 0.481 & 0.000 & 0.000 & 1.000 \\
Self-Employed & 8,777,905 & 0.241 & 0.428 & 0.000 & 0.000 & 0.000 \\
Agriculture & 8,478,755 & 0.113 & 0.317 & 0.000 & 0.000 & 0.000 \\
Industry & 8,478,755 & 0.405 & 0.491 & 0.000 & 0.000 & 1.000 \\
Trade & 8,478,755 & 0.266 & 0.442 & 0.000 & 0.000 & 1.000 \\
Administration & 8,478,755 & 0.175 & 0.380 & 0.000 & 0.000 & 0.000 \\
\hline
\multicolumn{7}{l}{\textit{Panel B: Municipal Data (1933)}} \\
\hline
NSDAP Share & 66,019 & 0.030 & 0.053 & 0.000 & 0.003 & 0.036 \\
SS Share & 66,019 & 0.001 & 0.014 & 0.000 & 0.000 & 0.000 \\
Population & 66,019 & 953.783 & 17808.060 & 143.000 & 291.000 & 577.000 \\
Jewish Share & 66,019 & 0.005 & 0.052 & 0.000 & 0.000 & 0.000 \\
Protestant Prob. & 66,019 & 0.690 & 0.365 & 0.515 & 0.903 & 0.957 \\
Protestant Prob. (NSDAP) & 66,019 & 0.692 & 0.355 & 0.524 & 0.892 & 0.955 \\
Female Share & 66,019 & 0.506 & 0.009 & 0.500 & 0.505 & 0.512 \\
Female Share (NSDAP) & 66,019 & 0.044 & 0.106 & 0.000 & 0.000 & 0.067 \\
Agriculture (Census) & 66,019 & 0.545 & 0.184 & 0.411 & 0.585 & 0.689 \\
Industry (Census) & 66,019 & 0.277 & 0.151 & 0.160 & 0.228 & 0.375 \\
Trade (Census) & 66,019 & 0.092 & 0.046 & 0.062 & 0.085 & 0.106 \\
Administration (Census) & 66,019 & 0.032 & 0.019 & 0.022 & 0.027 & 0.035 \\
Agriculture (NSDAP) & 66,019 & 0.052 & 0.116 & 0.000 & 0.000 & 0.083 \\
Industry (NSDAP) & 66,019 & 0.193 & 0.252 & 0.000 & 0.000 & 0.370 \\
Trade (NSDAP) & 66,019 & 0.112 & 0.179 & 0.000 & 0.000 & 0.208 \\
Administration (NSDAP) & 66,019 & 0.080 & 0.148 & 0.000 & 0.000 & 0.143 \\
Urban & 66,019 & 0.050 & 0.217 & 0.000 & 0.000 & 0.000 \\
East of Elbe & 66,019 & 0.571 & 0.495 & 0.000 & 1.000 & 1.000 \\
\hline
\multicolumn{7}{l}{\textit{Panel C: County Data (1933)}} \\
\hline
NSDAP Share & 966 & 0.064 & 0.268 & 0.031 & 0.047 & 0.063 \\
SS Share & 966 & 0.000 & 0.002 & 0.000 & 0.000 & 0.000 \\
Population & 966 & 65184.039 & 141470.565 & 30884.250 & 45716.500 & 69608.750 \\
Jewish Share & 966 & 0.005 & 0.006 & 0.001 & 0.003 & 0.007 \\
Protestant Prob. & 955 & 0.600 & 0.365 & 0.183 & 0.781 & 0.928 \\
Protestant Prob. (NSDAP) & 966 & 0.607 & 0.329 & 0.269 & 0.745 & 0.894 \\
Female Share & 926 & 0.510 & 0.014 & 0.502 & 0.509 & 0.517 \\
Female Share (NSDAP) & 966 & 0.084 & 0.018 & 0.081 & 0.085 & 0.089 \\
Agriculture (Census) & 966 & 0.429 & 0.256 & 0.193 & 0.483 & 0.649 \\
Industry (Census) & 966 & 0.347 & 0.179 & 0.187 & 0.318 & 0.470 \\
Trade (Census) & 966 & 0.121 & 0.079 & 0.067 & 0.095 & 0.149 \\
Administration (Census) & 966 & 0.042 & 0.035 & 0.023 & 0.030 & 0.046 \\
Agriculture (NSDAP) & 966 & 0.099 & 0.020 & 0.096 & 0.100 & 0.105 \\
Industry (NSDAP) & 966 & 0.363 & 0.060 & 0.361 & 0.369 & 0.375 \\
Trade (NSDAP) & 966 & 0.212 & 0.039 & 0.210 & 0.216 & 0.223 \\
Administration (NSDAP) & 966 & 0.150 & 0.029 & 0.148 & 0.154 & 0.159 \\
\hline\hline
\end{tabular}

}%
\begin{tablenotes}
\item \scriptsize{\textbf{Note.} Panel~A reports individual-level characteristics of NSDAP members. Panels~B and~C report municipality- and county-level data for 1933. Variables labeled ``(NSDAP)'' are computed from individual membership records aggregated to the respective geographic level. ``(Census)'' variables are employment shares from the 1925 \textit{Berufsz\"ahlung}. Protestant Prob.\ is the municipality-level Protestant share, constructed by combining 1925 census county shares with municipal-level church counts from the Meyers Gazetteer. Protestant Prob.~(NSDAP) is the mean of the individual-level Bayesian name-based Protestant probability across NSDAP members, which additionally uses name-specific likelihood ratios estimated from 1.6 million parish register records (see Section~\ref{sec:data} and Appendix~\ref{sec:appendix_protestant}). Female Share is the female share of \textit{Berufszugeh\"orige} (employed plus dependents) from 1933 census data. Female Share~(NSDAP) is the share of female members inferred from first names.}
\end{tablenotes}
\end{threeparttable}
\label{table:summary}
\end{small}
\end{table}

\subsection{Data Source}
\label{app:data_individual}

The NSDAP membership card files are held by the U.S. National Archives and Records Administration (NARA) as Record Group 242. The collection comprises two complementary card indexes (\textit{Karteien}):

\begin{enumerate}[nosep]
    \item \textbf{Zentralkartei} (Series A3340-MFKL). The central membership file maintained by the NSDAP's national membership office (\textit{Reichskartei}) in Munich. It contains 9,539,039 digitized microfilm images organized into letter groups A through T, with partial losses within Fa--G and Ka--O; letters U--Z are not present in this series.
    \item \textbf{Ortsgruppenkartei} (Series A3340-MFOK). The local group membership file, organized by \textit{Ortsgruppe}. It contains 6,723,785 digitized images across all 26 letter groups (A--Z).
\end{enumerate}

\noindent Together, the two series comprise \textbf{16,262,824} digitized microfilm images across 5,442 microfilm rolls. Each card records a member's name, birth date, occupation, membership number, and organizational affiliation.

Within each letter group, images are arranged on microfilm rolls (75--188 rolls per letter, approximately 2,500--3,500 pages per roll) in uncompressed TIFF format. Approximately 48\% of pages contain individual membership records; the remainder are blank frames, microfilm headers, or administrative sheets.

\subsection{Extraction Pipeline}

We extract structured data from each card image using Google's Gemini vision-language model (\texttt{gemini-3.1-flash-lite-preview}) via the Generative Language API. The pipeline operates in two stages. In the first stage, each TIFF image is downloaded from NARA's S3 storage, converted to RGB JPEG (quality~90) in memory, and uploaded to a Google Cloud Storage bucket. This conversion reduces per-image size from approximately 11\,MB to 1.7\,MB while retaining sufficient resolution for handwriting recognition. The conversion and upload run on Google Compute Engine virtual machines, processing disjoint letter groups in parallel.

In the second stage, each JPEG is submitted to the Gemini API with the extraction prompt shown in Listing~\ref{lst:prompt}. The API returns structured JSON conforming to a fixed schema (Listing~\ref{lst:schema}). Concurrent requests are managed by an asynchronous pipeline that logs all extractions for reproducibility. The prompt (Listing~\ref{lst:prompt}) instructs the model to transcribe all visible text, preserve original German spelling, and annotate illegible or struck-through entries. Disambiguation rules separate the occupation field (\textit{Beruf/Stand}) from marital status (\textit{Familienstand}).

\begin{lstlisting}[caption={Extraction prompt submitted with each card image.},label={lst:prompt},float=htbp]
Source: 1925-1945 NSDAP membership card (Mitgliedskarte).
Script: typeset and handwritten (Suetterlin/Kurrent), sometimes Fraktur.

First parse all visible text on the card. Then fill JSON fields from the
parsed text only. Do not infer missing values. Use empty string for
absent fields. Mark illegible text as [unclear: best guess]. Wrap
struck-through text in [crossed-out: text]. If multiple cards for
different persons on one scan, return one object per person. Layout
varies, some cards are just continuation pages with little information,
most have no portraits.

Format dates dd.mm.yyyy. If only year visible, use yyyy. If month+year,
use mm.yyyy.
beruf_stand: Do NOT confuse with Familienstand.
geborene: always a surname, never an occupation.
\end{lstlisting}

\begin{lstlisting}[caption={JSON response schema (field names and types).},label={lst:schema},float=htbp]
page_id              string   Unique identifier
nachname             string   Surname
vorname              string   Given name(s)
title_honorific      string   e.g. Dr., Prof., Ing., Dipl.
beruf_stand          string   Occupation (Beruf/Stand)
geborene             string   Maiden name
verehelichte         string   Married name
geb_datum            string   Date of birth (dd.mm.yyyy)
geb_ort              string   Place of birth
nsdap_mitgl_nr       string   Membership number (4-7 digits)
aufn_eingetr_eintritt string  Admission/entry date (dd.mm.yyyy)
aufn_beantragt       string   Admission requested (dd.mm.yyyy)
wiederaufn_beantragt string   Re-admission requested (dd.mm.yyyy)
wiederaufn_genehmigt string   Re-admission approved (dd.mm.yyyy)
austritt             string   Exit date (dd.mm.yyyy)
geloescht            string   Membership deleted (dd.mm.yyyy)
ausschluss           string   Expulsion (dd.mm.yyyy)
aufgehoben           string   Expulsion annulled (dd.mm.yyyy)
gestrichen_wegen     string   Reason for deletion
zurueckgenommen      string   Retracted (dd.mm.yyyy)
abgang_von_wehrmacht string   Departure to Wehrmacht (dd.mm.yyyy)
zugang_von_wehrmacht string   Return from Wehrmacht (dd.mm.yyyy)
gestorben            string   Death date (dd.mm.yyyy)
wohnort              string   Place of residence
strasse              string   Street
hausnummer           integer  House number
ortsgr               string   Local group (Ortsgruppe)
gau                  string   Regional district (Gau)
wohnung_weitere      array    Further addresses [{wohnort, strasse,
                              hausnummer, ortsgr, gau, date}]
move_to              string   New location after moving
move_date            string   Move date (dd.mm.yyyy)
bemerkungen          string   All text not fitting another field
dienststelle         string   Office/post
membership_other_ns  string   e.g. SS, SA, NSV, HJ
verwarnung           string   e.g. Verwarnung, Schwarze Liste, OPG
family_status        string   ledig, verheiratet, verwitwet, geschieden
family_status_children integer 1 if children noted
handwritten          integer  1 if card is handwritten
portrait             string   Semicolon-delimited descriptors
vormerkkarte         integer  1 if preliminary card
not_legible          integer  1 if card is not legible
\end{lstlisting}

\subsection{Variables Extracted}
\label{sec:variables}

Table~\ref{tab:variables} describes the fields extracted from each card image. Fill rates are conditional on the page containing an identifiable membership record (non-blank scans).

\begin{longtable}{@{} l l p{6cm} r @{}}
\caption{Variables extracted from NSDAP membership cards.} \label{tab:variables} \\
\toprule
Field & Type & Description & Fill (\%) \\
\midrule
\endfirsthead
\multicolumn{4}{@{}l}{\small\itshape Table~\ref{tab:variables} continued} \\
\toprule
Field & Type & Description & Fill (\%) \\
\midrule
\endhead
\bottomrule
\multicolumn{4}{r}{\small\itshape continued on next page} \\
\endfoot
\bottomrule
\multicolumn{4}{@{}p{\textwidth}@{}}{\footnotesize \textit{Notes:} Fill rates are conditional on pages containing identifiable membership records (non-blank scans). Administrative fields (exit, expulsion, death) have low fill rates because most members did not experience these events during the observation period.} \\
\endlastfoot
\texttt{page\_id} & ID & Unique NARA identifier (e.g., A0004-02870) & 100.0 \\
\texttt{name} & Text & Surname (\textit{Familienname}) & 100.0 \\
\texttt{vorname} & Text & Given name(s) (\textit{Vorname}) & 93.5 \\
\texttt{geborene} & Text & Maiden name, married women only & 3.9 \\
\texttt{geb\_datum} & Date & Date of birth & 92.1 \\
\texttt{geb\_ort} & Text & Place of birth & 86.6 \\
\texttt{stand} & Text & Occupation (\textit{Beruf/Stand}) & 87.0 \\
\texttt{mitgl\_nr} & Number & NSDAP membership number & 92.3 \\
\texttt{aufnahme\_datum} & Date & Date of admission/entry to the NSDAP & 89.3 \\
\texttt{aufnahme\_beantragt} & Date & Date admission was requested & 45.5 \\
\texttt{wiederaufnahme} & Date & Re-admission request or grant & 2.0 \\
\texttt{austritt} & Date/Text & Exit from the party & 2.7 \\
\texttt{geloescht} & Date/Text & Membership deleted & 0.1 \\
\texttt{ausschluss} & Date/Text & Expulsion & 0.3 \\
\texttt{aufgehoben} & Date/Text & Expulsion annulled & $<$0.1 \\
\texttt{gestrichen\_wegen} & Text & Reason for deletion & 1.4 \\
\texttt{zurueckgenommen} & Date/Text & Membership retracted & 0.2 \\
\texttt{abgang\_wehrmacht} & Date/Text & Departure to Wehrmacht & $<$0.1 \\
\texttt{zugang\_von} & Text & Transfer from Wehrmacht & $<$0.1 \\
\texttt{dienststelle} & Text & NS duty station (\textit{Dienststelle}) & 0.8 \\
\texttt{gestorben} & Date/Text & Death recorded on card & 0.9 \\
\texttt{bemerkungen} & Text & Unstructured remarks field & 39.7 \\
\texttt{wohnung} & Text & Address (\textit{Wohnung}) & 88.5 \\
\texttt{ortsgr} & Text & Local group (\textit{Ortsgruppe}) & 82.8 \\
\texttt{gau} & Text & Regional district (\textit{Gau}) & 81.9 \\
\texttt{wohnung\_weitere} & Text & Additional addresses, semicolon-delimited & 22.9 \\

\texttt{card\_type}        & Cat.      & Card format: ``small'' or ``large''                & 100.0 \\
\end{longtable}

\subsection{Extraction Challenges}
\label{sec:challenges}

Three features of the source material introduce measurement error.

\paragraph{Handwriting and layout.}
Thousands of local officials completed cards by hand between 1925 and 1945, producing wide variation in legibility. The model marks text it cannot decipher as \texttt{[unclear]}; approximately 2--5\% of fields per card contain this annotation. Two card formats appear in the collection: \textit{small cards} (index-card sized, single page) and \textit{large cards} (A5, two to three consecutive microfilm pages).

\paragraph{Field disambiguation.}
The occupation field (\textit{Beruf}) appears in varying positions across card formats. Dates appear in inconsistent formats and require standardization. Membership numbers occasionally include thousand separators or textual prefixes.

\subsection{Post-Processing}
\label{sec:postprocessing}

The raw extraction output undergoes five post-processing steps.

\begin{enumerate}[nosep,label=(\roman*)]
\item \textit{Multi-page linking.} Continuation pages inherit the name and membership number from the most recent preceding front page within the same roll.
\item \textit{Date standardization.} Dates are converted to ISO~8601 format, with regime-period dates validated to fall within 1919--1945 and birth dates within 1840--1930.
\item \textit{Membership number normalization.} Thousand separators and textual prefixes are stripped.
\item \textit{Abbreviation expansion.} Abbreviated location fields are matched to full names elsewhere on the same card.
\item \textit{Bemerkungen parsing.} Structured information (e.g., expulsions) is extracted from the unstructured remarks field via pattern matching.
\end{enumerate}

\subsection{Occupation Classification}
\label{sec:occupation}

The occupation field (\textit{Beruf/Stand}) records free-text entries in German, often abbreviated (e.g., ``Kfm.'' for \textit{Kaufmann}, ``Schlosserges.'' for \textit{Schlossergeselle}). We classify each unique occupation string into the categories of the 1925 occupational census (\textit{Berufsz\"ahlung}), following the classification of the Statistisches Reichsamt (StDR Bde.~402--405). The classification yields two variables: \textit{Stellung im Beruf} (occupational status: self-employed, salaried employee, wage worker, civil servant, or family worker) and industry sector. A language model maps each raw string to the census categories, preserving the original German occupational taxonomy. We also retain an alternative ISCO/NACE classification. The main text reports the workplace specification at the municipality-by-industry-by-occupational-status level. If the model was not able to infer either industry or occupation, we treat these observations as unclassifiable.

\subsection{Derived Individual Variables}
\label{sec:derived_vars}

\paragraph{Gender.}
We classify each member's gender from their first name in two steps. First, we match against a curated German first-name dictionary with known gender assignments. Second, for unmatched names, we apply a suffix heuristic: names ending in \textit{-a}, \textit{-ia}, \textit{-ie}, \textit{-ina}, \textit{-ine}, or \textit{-ette} are classified as female.

\paragraph{SS membership.}
We identify SS members from two sources. First, we flag all individuals whose NSDAP card text mentions an ``SS'' affiliation in the organizational membership or remarks fields. Second, we link our records to approximately 50,000 individuals named in the \textit{SS-Verordnungsblatt} \citep{dwsxip_ss_verordnungsblatt}, matching on membership number where available and on name and date of birth otherwise. The external source provides SS entry date, rank, and promotion history.

\paragraph{Academic title and nobility.}
We detect academic titles and aristocratic status via pattern matching on the name and profession fields. Doctoral degrees are flagged by matching ``Dr.'' or ``Doktor.'' Nobility is identified by matching over 25 German aristocratic prefixes and titles, including \textit{von}, \textit{zu}, \textit{Freiherr}, \textit{Graf}, and their feminine variants.

\paragraph{Residential mobility.}
Membership cards record address changes in the \textit{Wohnung weitere} field, a list of further addresses with dates. We extract the first residential move: destination municipality, street, and date. When this field is empty, we use the separate move-to and move-date fields recorded elsewhere on the card.

\paragraph{Deduplication.}
Members may appear on multiple cards across the \textit{Zentralkartei} and \textit{Ortsgruppenkartei}. We deduplicate on a composite key: membership number (restricted to numbers with six or more digits), first name, surname, and birth year. Cards sharing the same composite key are merged into a single record, retaining the most complete entry.

\begin{figure}[htpb!]
\caption{Sample Membership Card: Machine-Printed}
\centering
\fbox{\parbox{0.8\textwidth}{\includegraphics[width=0.8\textwidth]{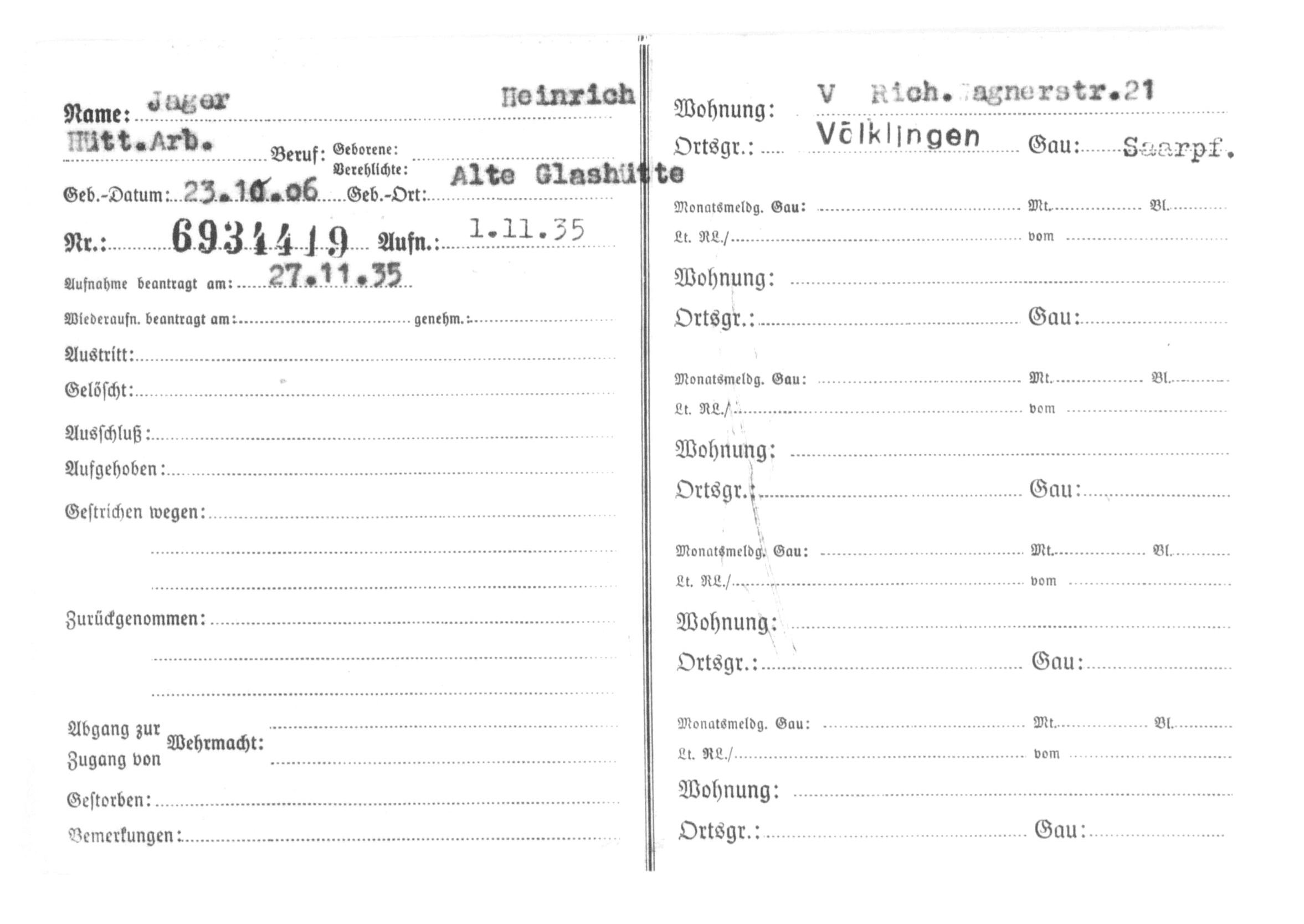}}}
\label{fig:card_printed}
\\[4pt]
\justifying
\scriptsize{\textbf{Note.} Machine-printed NSDAP membership card from the \textit{Zentralkartei}. Fields are filled by typewriter.}
\end{figure}

\begin{figure}[htpb!]
\caption{Sample Membership Card: Handwritten}
\centering
\fbox{\parbox{0.5\textwidth}{\includegraphics[width=0.5\textwidth]{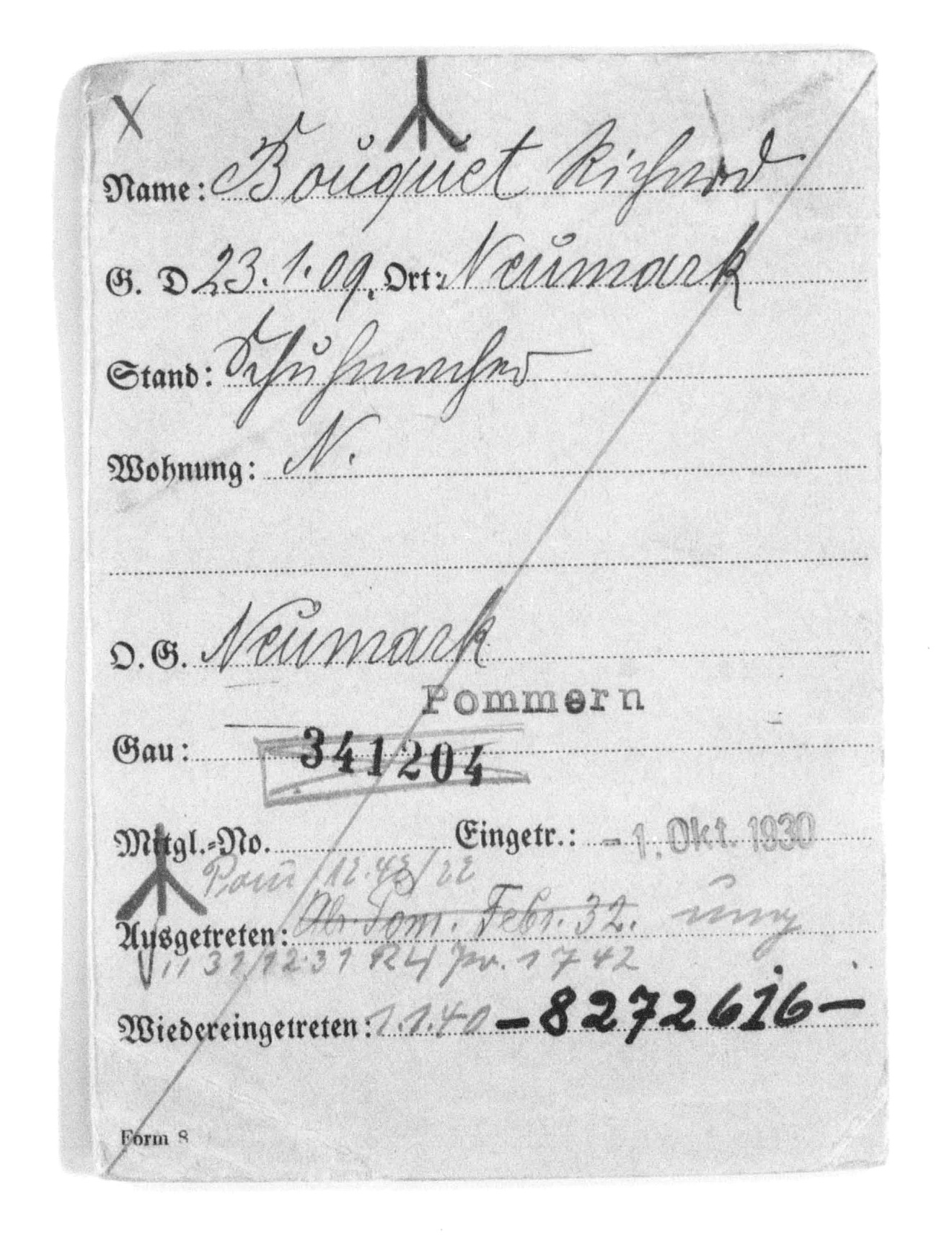}}}
\label{fig:card_handwritten}
\\[4pt]
\justifying
\scriptsize{\textbf{Note.} Handwritten NSDAP membership card from the \textit{Zentralkartei}. Fields are completed in cursive by a local party official.}
\end{figure}

\begin{figure}[htpb!]
\caption{Sample Membership Card: Portrait With NS Symbols}
\centering
\fbox{\parbox{0.8\textwidth}{\includegraphics[width=0.8\textwidth]{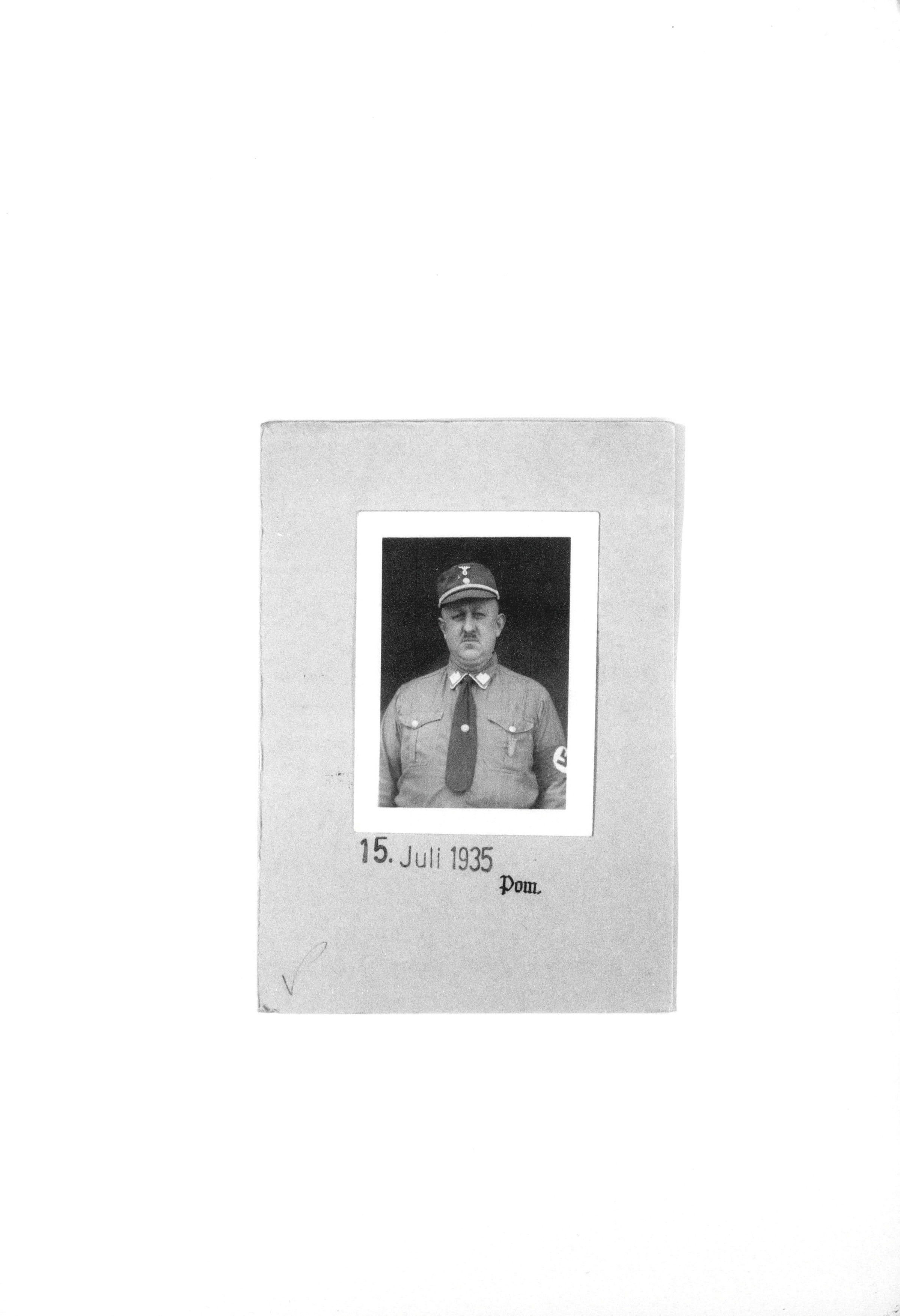}}}
\label{fig:card_portrait}
\\[4pt]
\justifying
\scriptsize{\textbf{Note.} Portrait in NSDAP membership card with NS symbols.}
\end{figure}

\begin{figure}[htpb!]
\caption{Sample Membership Card: Portrait With NS Symbols}
\centering
\fbox{\parbox{0.8\textwidth}{\includegraphics[width=0.8\textwidth]{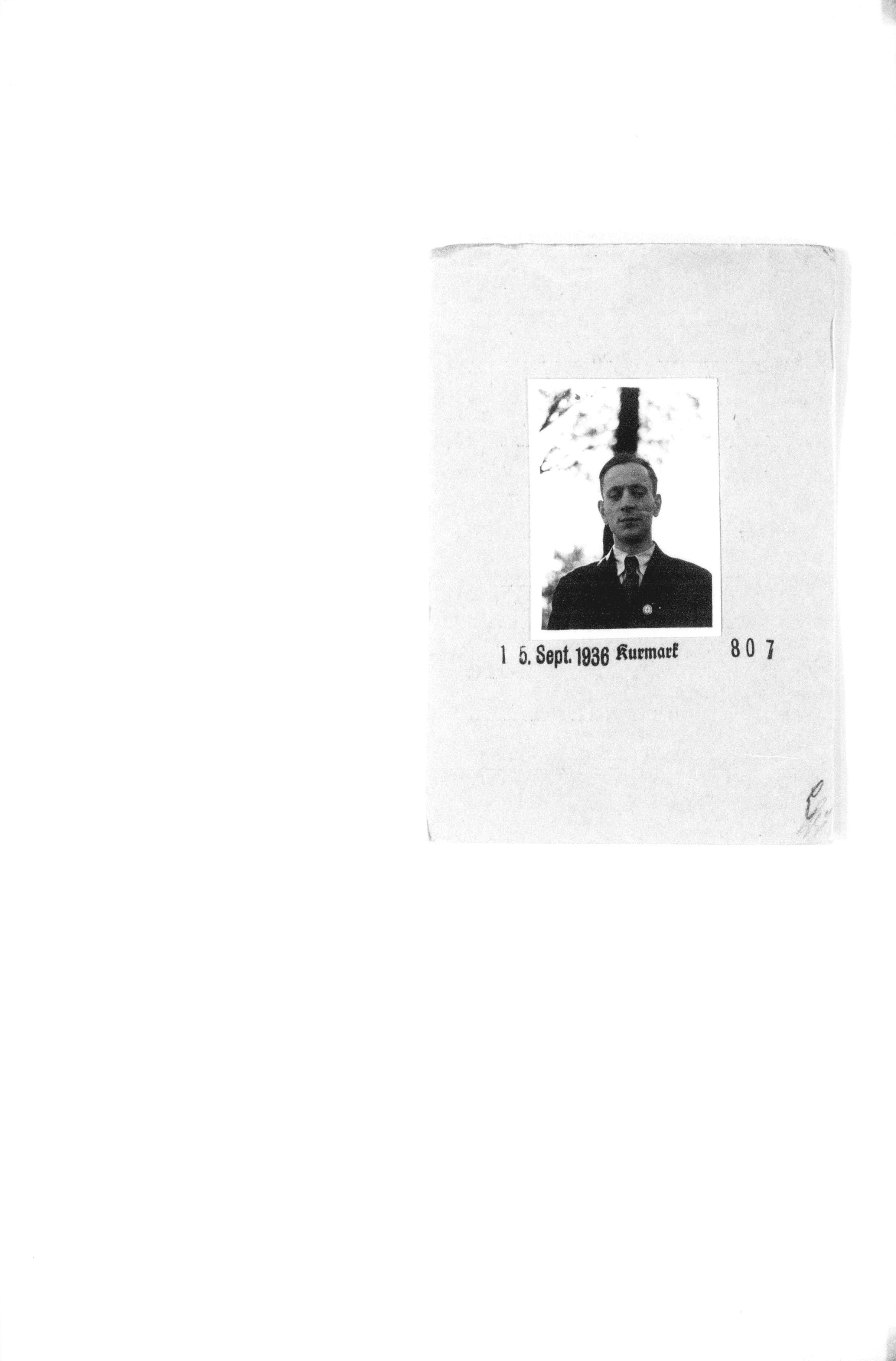}}}
\label{fig:card_portrait_2}
\\[4pt]
\justifying
\scriptsize{\textbf{Note.} Portrait in NSDAP membership card with NS symbols.}
\end{figure}

\begin{figure}[htpb!]
\caption{Sample Membership Card: Portrait Without NS Symbols}
\centering
\fbox{\parbox{0.8\textwidth}{\includegraphics[width=0.8\textwidth]{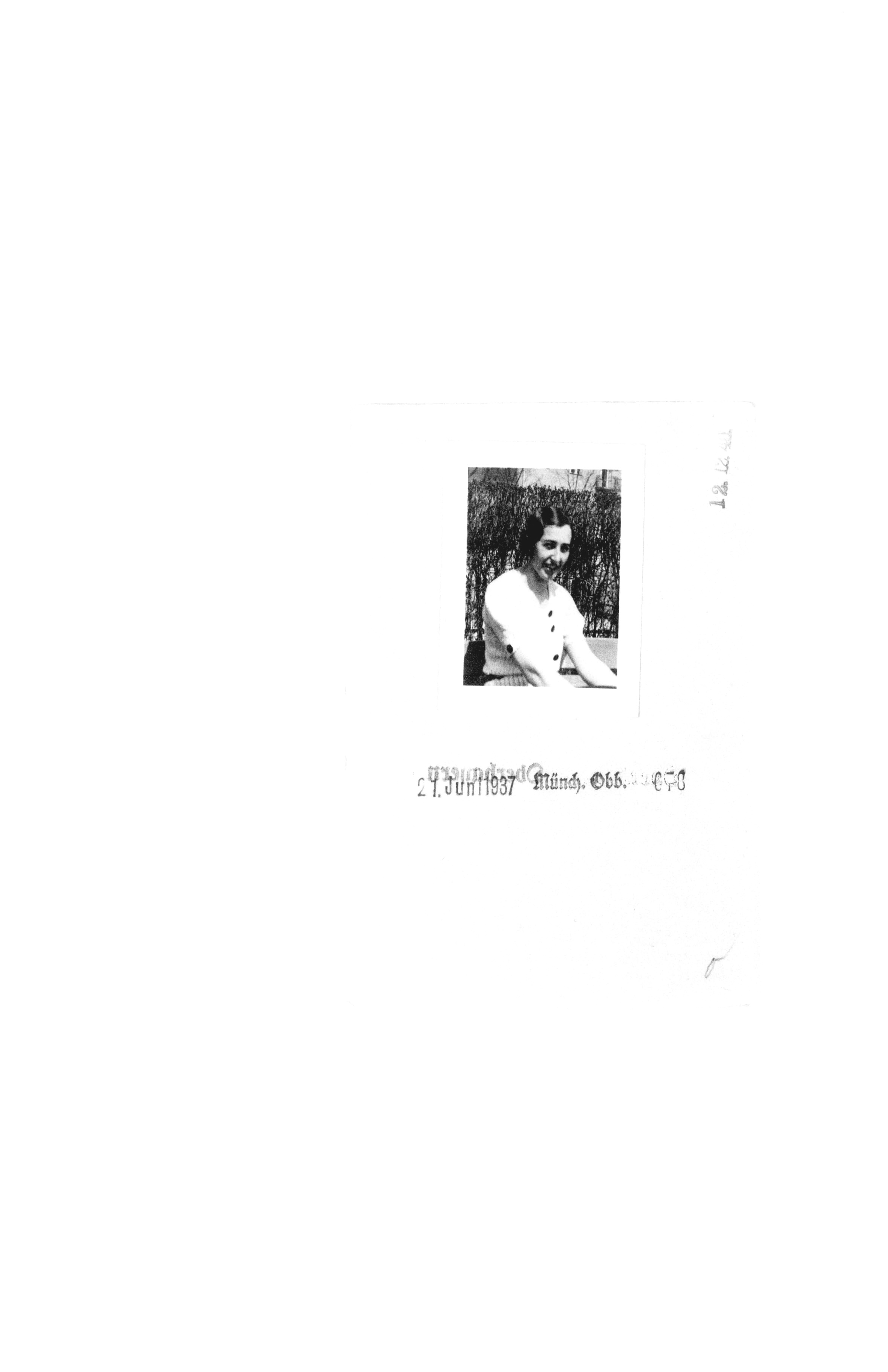}}}
\label{fig:card_portrait_3}
\\[4pt]
\justifying
\scriptsize{\textbf{Note.} Portrait in NSDAP membership card without NS symbols.}
\end{figure}

Figures~\ref{fig:card_printed} and~\ref{fig:card_handwritten} show sample membership cards in the two main formats. Figures~\ref{fig:card_portrait} and~\ref{fig:card_portrait_2} show sample portraits with NS insignia. Figure~\ref{fig:card_portrait_3} shows a sample portrait without NS insignia. 

\paragraph{Implications for inference.}
The above extraction challenges may introduce measurement error that might not be classical \citep{carlson2025unifying}. If legibility correlates with card age, region, or member prominence, estimates may exhibit non-random attenuation bias. We address this by (i)~using the \texttt{[unclear]} and \texttt{[crossed-out: ...]} annotations to flag uncertain observations, (ii)~validating key variables against external sources, and (iii)~restricting to machine-printed cards or controlling for extraction difficulty. 

\subsection{Reconstruction of Municipal Boundaries}
\label{sec:municipal_boundaries}

No digitized shapefile of municipal boundaries exists for the German Empire. County-level (\textit{Kreis}) boundaries are available from the Census Mosaic project, but analyses at the municipality level require finer geographic units. We reconstruct approximate municipal polygons using Voronoi tessellation within county borders.

\paragraph{Geocoding.} We assign geographic coordinates to all municipalities listed in the 1910 \textit{Gemeindeverzeichnis} classified as \textit{Landgemeinde} (rural municipality) or \textit{Stadt} (town). Coordinates are drawn from three sources in priority order. For cities, we use a curated database of historical city locations. For the remaining municipalities, we match place names to the \textit{Meyers Orts- und Verkehrslexikon des Deutschen Reichs}, a historical gazetteer that records coordinates for approximately 130,000 German localities. Matching uses text embeddings and cosine similarity. We condition our matches to be within the same county, with population as a secondary criterion to resolve ties. For the remaining unmatched municipalities, we geocode using a language model, providing the municipality name, county, and state as context. To detect coordinate errors, we compare each municipality's assigned location to the median coordinate of its county and district, flagging outliers beyond a multiple of the median absolute deviation. Flagged coordinates are replaced by further rounds of geocoding until all flagged coordinates are resolved. The procedure yields over 65,000 geocoded point locations with population counts.

\paragraph{Voronoi tessellation.} The tessellation proceeds in three steps. First, we assign each municipality point to its enclosing county using the 1918 county boundaries from the Census Mosaic GIS collection. Second, within each county polygon, we compute a Voronoi tessellation of the municipality points and clip each cell to the county boundary. Third, we validate that (i)~every point falls within exactly one municipal polygon, (ii)~county borders are preserved as polygon edges, and (iii)~\textit{Stadtkreise} (urban counties containing a single city) coincide with the county boundary.

The resulting tessellation partitions the territory of the German Empire into over 65,000 non-overlapping municipal polygons. The Voronoi cells approximate true boundaries well where nearby settlements have similar population shares. For our purposes, the relevant requirement is that each geocoded address is assigned to the correct municipality, not that the polygon boundary is cartographically exact.

\paragraph{Census and election data.} The census and election returns compiled by \citet{falter1991hitlers} report demographic and voting outcomes at the county (\textit{Kreis}) level, but disaggregate municipalities with more than 2,000 inhabitants as separate observations within each county. We leverage this structure to maintain maximum geographic resolution. Using the reference coordinates provided with the election returns, we spatially assign each Voronoi polygon to its county in the Falter data via nearest-neighbor matching. Within each county, we then link individually reported municipalities to their nearest Voronoi polygon; the remaining polygons receive the residual county aggregate. This procedure preserves sub-county variation wherever the source data provide it, rather than collapsing all municipalities to a single county-level observation.

\paragraph{Place-name matching.} Each membership card records up to three place names (residence, birthplace, and \textit{Ortsgruppe}) as free text. We link these to the geocoded municipalities using text embeddings and cosine similarity to retrieve candidate matches. Because many German place names recur across regions (e.g., multiple municipalities named ``Neustadt''), embedding similarity alone is insufficient.

We resolve ambiguities in three steps. First, we restrict candidates to the same \textit{Gau} as the card's recorded residence. Second, we rank by population similarity: for each candidate municipality~$m$ in \textit{Gau}~$g$, we compare the share of cards mentioning the associated place name ($\hat{s}_m = n_m / n_g$) to the municipality's 1910 census population share ($s_m = \mathrm{pop}_m / \mathrm{pop}_g$), scoring by $1 - |\hat{s}_m - s_m|$. This penalizes assigning a frequent place name to a small hamlet or a rare name to a large city. Third, embedding cosine similarity breaks remaining ties.

\subsection{Protestant Classification}
\label{sec:appendix_protestant}

We estimate each NSDAP member's Protestant probability rather than a binary indicator, combining three layers of progressively finer information.

\paragraph{County prior.} The 1925 population census records Protestant and Catholic population counts at the county level (\textit{Kreis}). These shares serve as the prior probability that a randomly drawn resident is Protestant.

\paragraph{Municipal refinement.} The \textit{Meyers Orts- und Verkehrslexikon} records the number of Protestant (\textit{evangelische Kirchen}) and Catholic (\textit{katholische Kirchen}) churches for each municipality. Within each county, we compute the population-weighted mean church ratio $\bar{R}_c = \sum_m \text{pop}_m \cdot r_m / \sum_m \text{pop}_m$, where $r_m = n^{\text{ev}}_m / (n^{\text{ev}}_m + n^{\text{kath}}_m)$. The municipal Protestant share is then $P_m = P_c \cdot r_m / \bar{R}_c$, clamped to $[0.01, 0.99]$. This allocates the county share across municipalities in proportion to local church composition. Municipalities without churches retain the county share.

\paragraph{Name signal.} We estimate name-specific likelihood ratios from 1.6 million records in digitized German parish registers \citep{verein2023ofb}, which directly observe both first name and religious denomination. For each first name $n$, we compute the Protestant share $\hat{p}_n$ and convert to a likelihood ratio $\Lambda_n = [\hat{p}_n / (1-\hat{p}_n)] \,/\, [\bar{p} / (1-\bar{p})]$, where $\bar{p}$ is the overall Protestant share in the parish register sample. This ratio is invariant to the sample's denominational composition. Table~\ref{tab:name_lr} reports selected examples. The final individual probability is $P(\text{Prot} \mid \text{name}, \text{municipality}) = \Lambda_n P_m \,/\, [\Lambda_n P_m + (1 - P_m)]$.

\begin{table}[h]
\centering
\caption{Name-Religion Association: Selected First Names}
\label{tab:name_lr}
\small
\resizebox{\textwidth}{!}{%
\begin{tabular}{llrr@{\hskip 18pt}llrr@{\hskip 18pt}llrr}
\toprule
\multicolumn{4}{c}{Protestant names} &
\multicolumn{4}{c}{Catholic names} &
\multicolumn{4}{c}{Ambiguous names} \\
\cmidrule(r){1-4} \cmidrule(r){5-8} \cmidrule(l){9-12}
Name & $N$ & \% Prot & $\Lambda$ &
Name & $N$ & \% Prot & $\Lambda$ &
Name & $N$ & \% Prot & $\Lambda$ \\
\midrule
Gottlieb  &   3{,}777 & 94.3 & 7.54 &
Alois     &   1{,}202 &  2.2 & 0.01 &
Karl      &  10{,}603 & 72.2 & 1.18 \\
Friedrich &  21{,}854 & 92.9 & 5.92 &
Ignaz     &     311   &  2.4 & 0.01 &
Martin    &   7{,}974 & 75.9 & 1.43 \\
Ernst     &   6{,}547 & 91.0 & 4.57 &
Josef     &   7{,}861 &  6.3 & 0.03 &
Anna      & 116{,}224 & 70.9 & 1.11 \\
Heinrich  &  30{,}243 & 83.8 & 2.34 &
Franz     &  18{,}672 & 19.1 & 0.11 &
          &           &      &      \\
Johann    & 154{,}338 & 81.9 & 2.06 &
Maria     &  59{,}869 & 53.9 & 0.53 &
          &           &      &      \\
Wilhelm   &  15{,}798 & 80.7 & 1.90 &
Elisabeth &  14{,}520 & 54.8 & 0.55 &
          &           &      &      \\
\bottomrule
\end{tabular}%
}
\vspace{4pt}
\justifying
\scriptsize{\textbf{Note.} Likelihood ratios ($\Lambda$) estimated from 1.6 million parish register records \citep{verein2023ofb} with direct observation of religious denomination. $\Lambda > 1$ indicates a name more common among Protestants; $\Lambda < 1$ more common among Catholics. $N$ is the number of observations with recorded denomination.}
\end{table}

\begin{figure}[htpb!]
\caption{Validation of Protestant Classification}
\centering
\includegraphics[width=0.55\textwidth]{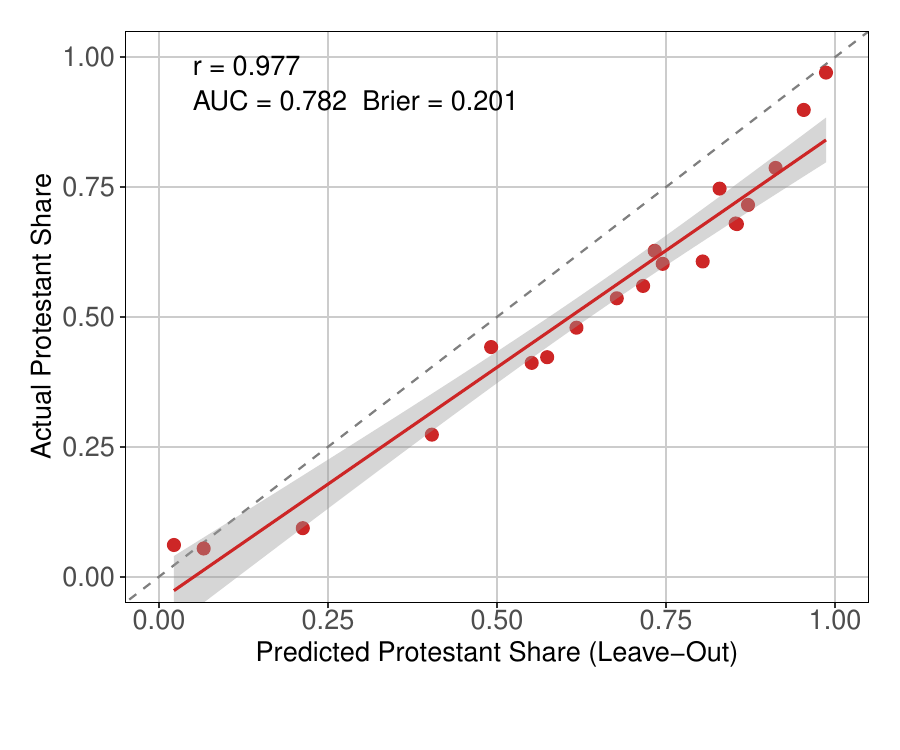}
\label{fig:protestant_leaveout}
\\[4pt]
\justifying
\scriptsize{\textbf{Note.} Twenty percent of parish register books are held out at the book level; name-specific Protestant shares are estimated from the remaining 80 percent with shrinkage ($K = 5$). Binned scatter of predicted versus actual Protestant share for 317,251 held-out individuals (20 equal-sized bins). The dashed line indicates perfect calibration.}
\end{figure}

Figure~\ref{fig:protestant_leaveout} validates the name component of the classification. We hold out 20 percent of parish register books entirely and estimate name-specific Protestant shares from the remaining 80 percent. For approximately 317,000 held-out individuals, the predicted Protestant share tracks the actual share closely ($r = 0.977$, AUC $= 0.782$, Brier score $= 0.201$).

\subsection{Jewish Population and Deportations}
\label{sec:appendix_jewish}

\paragraph{Jewish communities.} We compile historical Jewish community population counts from approximately 4,600 localities across Germany, drawing on the \textit{J\"udische Gemeinden} online archive \citep{juedischegemeinden}. We match locality names to our geocoded municipalities and aggregate counts to the municipal level. Where census observations are available for multiple years but not annually, we interpolate intervening years log-linearly between observed data points. These municipal-level counts enter the cross-sectional analysis as a control for historical Jewish population shares.

\paragraph{Deportations and victims.} We obtain individual-level records on Jewish residents, deportees, and emigrants from three sources: the List of Jewish Residents in Germany 1933--1945, the 1939 German Minority Census, and the German Federal Archives' memorial book (\textit{Gedenkbuch}) \citep{tracingthepast2018mappingthelives, bundesarchiv2012gedenkbuch}. Each record contains a residential address, which we geocode to a municipality via spatial intersection with our Voronoi polygons. We aggregate to municipal-level counts of Jewish residents, deportees, and emigrants. To measure individual-level proximity between NSDAP members and Jewish residents, we parse residential addresses into street name and house number, apply the same street-name normalization used for the membership cards, and match to party members at three levels: same municipality, same street, and same house number.

\subsection{Motives for Joining (Abel Reports)}
\label{sec:appendix_abel}

\paragraph{Data source.} We draw on 580 autobiographical reports collected in 1934 by sociologist Theodore Abel as part of a prize competition organized by the NSDAP, in which members were invited to explain why they had joined the movement \citep{abel1938why}. Reports average 2--5 pages. Participants self-selected into the competition, so the sample is not random and is skewed toward early, ideologically committed joiners. We therefore treat the evidence as qualitative corroboration of the heterogeneity of motives, not as population inference.

\paragraph{Coding.} We code each report with a large language model (GPT-4o-mini) using a structured prompt that assigns binary indicators for five causal motives: economic hardship, antisemitism, anti-communism, national renewal/order, and social belonging. The prompt codes a motive as one only when the text explicitly links that factor to the decision to join the party. Background characteristics, general ideological statements, and post-entry activities are excluded. Reports shorter than 80 words are dropped. Table~\ref{tab:abel_quotes} shows one illustrative German-language passage per category. Figure~\ref{fig:abel_motives} in the main text reports the resulting distribution of motives.

\begin{table}[htpb!]
\caption{Example Causal Quotes by Motive Category}
\label{tab:abel_quotes}
\centering
\footnotesize
\begin{tabular}{p{3.4cm} p{10cm}}
\toprule
Category & Quote \\
\midrule
Economic hardship      & \textit{``Ich wurde nach 14 j\"ahriger ununterbrochenen Arbeitszeit arbeits- und brotlos gemacht.''} \\
Antisemitism           & \textit{``Der Jude sa{\ss} an der Regierung und verschacherte unser Vaterland mit allen ihm zu Gebote stehenden Mitteln.''} \\
Anti-communism         & \textit{``Die Gefahr des Bolschewismus veranlasste mich zum Eintritt.''} \\
National renewal/order & \textit{``So ist das Letzte aus langem Erleben die Sehnsucht nach einem freien, starken Deutschland.''} \\
Social belonging       & \textit{``Die Kameradschaft zog mich an.''} \\
\bottomrule
\end{tabular}
\\[4pt]
\justifying
\scriptsize{\textbf{Note.} One illustrative verbatim passage per category, drawn from the coded corpus.}
\end{table}

\subsection{Validation Against the Falter/Kater Membership File}
\label{sec:appendix_falter}

Prior quantitative work on the social composition of the NSDAP relies on the 0.5\% stratified-cluster sample of membership cards compiled by \citet{kater1983nazi} and \citet{falter1991hitlers} and released as the NSDAP Membership File \citep{falter2022mitglieder}. The file contains 43{,}731 records, drawn proportionally within county strata. We apply our coding pipeline to these records and to the full population of membership cards, and compare the two sources on the same set of occupational and gender shares. We then use the full population to characterize what the 0.5\% sample can and cannot identify in local geographic comparisons of the party relative to its local population base.

\begin{figure}[htpb!]
\caption{Validating our Coding and Identification at the 0.5\% Membership Sample}
\centering
\includegraphics[width=0.75\textwidth]{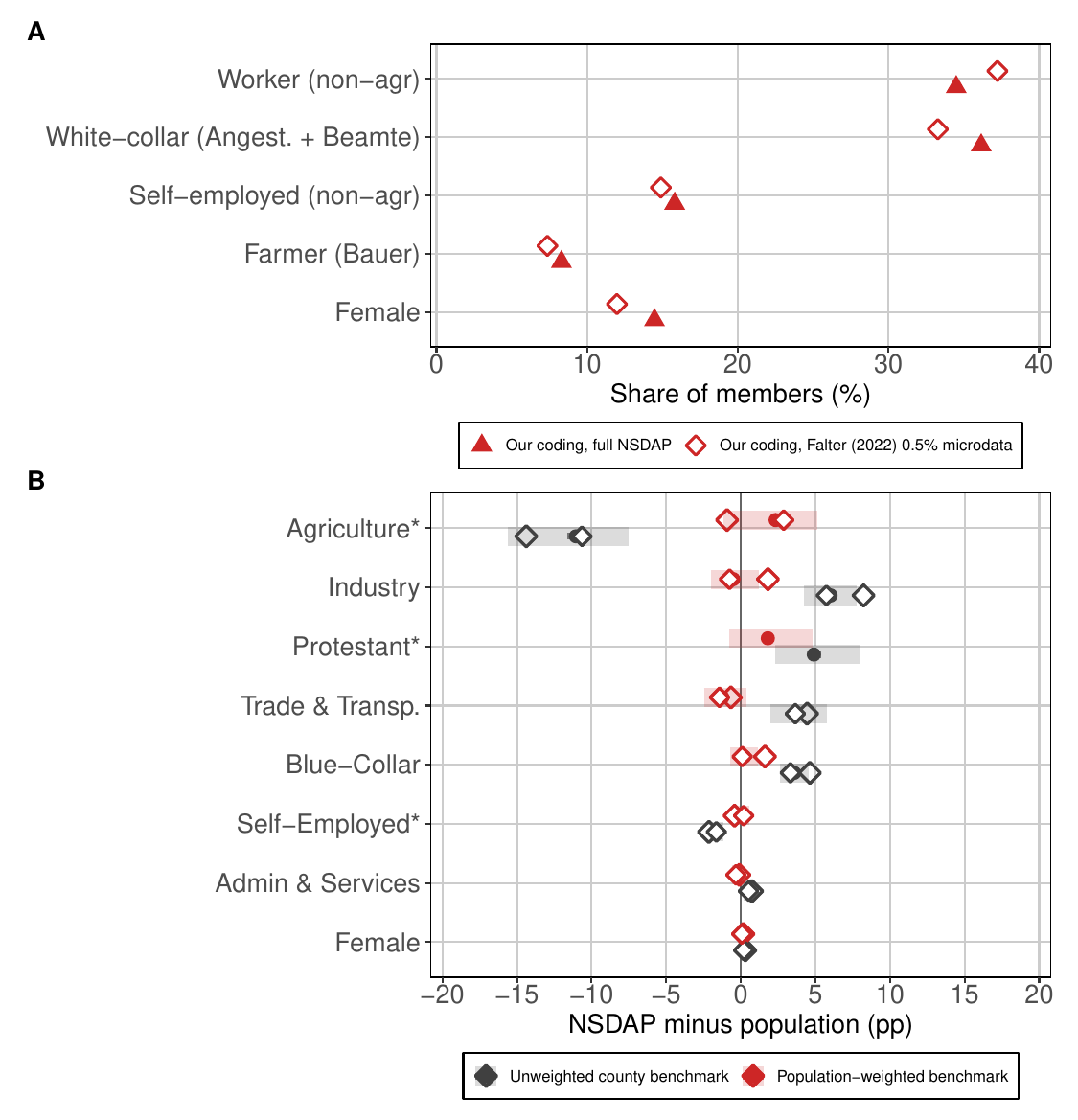}
\label{fig:falter_representativeness}
\\[4pt]
\justifying
\scriptsize{\textbf{Note.} Panel~\textbf{A}: marginal occupational and gender shares on the full 1925--1945 member universe. Red triangles are our coding applied to the full population of NSDAP cards; open red diamonds are the same coding applied to the \citet{falter2022mitglieder} 0.5\% microdata. Panel~\textbf{B}: between-county gap $\Delta_k^{g,\textup{between}}$ on the full member universe. Red = population-weighted estimator; black = unweighted cross-county estimator. Filled dots are full-population point estimates. Narrow bars are iid simple-random-sample 95\% bootstrap intervals at $n = 42{,}176$ (500 draws); wide bars are county-cluster 95\% bootstrap intervals that draw counties with replacement and draw members per county in proportion to the county's NSDAP count, accounting for the county-level clustering of the \citet{falter2022mitglieder} sample design. Open diamonds overlay \citet{falter2022mitglieder} records with raw \texttt{weight\_rel}; filled diamonds overlay the same records post-stratified so per-county weight sums equal the full NSDAP per-county count. Starred rows mark characteristics for which \citet{falter1991hitlers} reports NSDAP over-representation. See Appendix~\ref{sec:appendix_falter} for discussion.}
\end{figure}

Panel~A of Figure~\ref{fig:falter_representativeness} reports marginal occupational and gender shares on the full 1925--1945 member universe for two series: our coding applied to the full population of NSDAP cards (filled red triangles) and the same coding applied to the \citet{falter2022mitglieder} microdata (open red diamonds). Classification rules at the employment-status and industry-branch level in both series closely follow the 1925 \textit{Berufsz\"ahlung} categories (Appendix~\ref{sec:occupation}). The two series agree within one percentage point on non-agricultural self-employed ($-0.9$~pp) and farmers ($-0.9$~pp). The residual gaps are directional: relative to the full cards, the 0.5\% sample records more workers ($+2.7$~pp) and fewer white-collar members ($-2.9$~pp), and fewer women ($-2.5$~pp). The occupational discrepancies are likely due to residual uncertainty in the Arbeiter--Angestellte distinction. For the 1925--1932 entry cohort (``Alte K\"ampfer''), on which \citet{kater1983nazi} reports statistics, our pipeline reproduces his shares within a classification-judgement tolerance of on average 2 percentage points on all five categories.

Panel~B reports the between-county composition gap $\Delta_k^{g,\textup{between}}$ for each of the eight main-text characteristics under two aggregations of the within-county gap $g_i - p_i$ between NSDAP members and the local population. The population-weighted estimator (red) aggregates by NSDAP count: it asks whether the typical NSDAP member is more agrarian, Protestant, or self-employed than the population of their county. The unweighted cross-county estimator (black) treats each of Germany's roughly one thousand counties as one observation, counting a small rural county as much as Berlin: it asks how the within-county composition gap varies across county types. county size is heavy-tailed, so the two aggregations can disagree in sign. On the full 1925--1945 member universe the two estimators agree in sign on four characteristics and disagree on four: Agriculture ($+2.34$ vs.~$-11.00$~pp), Industry ($-0.51$ vs.~$+6.02$~pp), Self-Employed ($+0.16$ vs.~$-1.68$~pp), and Trade~\&~Transport ($-1.23$ vs.~$+3.72$~pp).

To characterize what a 0.5\% sample identifies, we draw 500 sub-samples of size $n = 42{,}176$ from the full member universe under two designs and report the 2.5th and 97.5th bootstrap percentiles. The narrow iid bars treat members as independent draws. The wide cluster bars resample counties with replacement and draw members per county in proportion to the county's NSDAP count, matching the county-level clustering of the \citet{falter2022mitglieder} release; at a 0.5\% sampling fraction a typical rural county yields two or three sampled cards, enough to rank counties but not to recover each county's internal composition. Under the clustered design the population-weighted interval covers zero for 8 of 8 characteristics, including every starred row; the unweighted interval excludes zero for 8 of 8. A 0.5\% county-clustered sample identifies the unweighted cross-county estimator and does not identify the population-weighted estimator.

The three starred rows in Panel~B --- Protestant, Agriculture, and Self-Employed --- mark characteristics for which \citet{falter1991hitlers} reports NSDAP over-representation in Protestant, agrarian, and Mittelstand communities. The population-weighted full-population estimates are small: Protestant $+1.81$~pp, Agriculture $+2.34$~pp, Self-Employed $+0.16$~pp; cluster intervals cover zero. The unweighted cross-county full-population estimates, which the 0.5\% sample identifies, are Protestant $+4.90$~pp, Agriculture $-11.00$~pp, and Self-Employed $-1.68$~pp. Prior over-representation claims --- \citet{falter1991hitlers}'s cross-county regressions of NSDAP penetration on covariates --- weight each county equally and belong to the family a 0.5\% cluster sample identifies. The membership-weighted aggregation, which weights each NSDAP member once, is a different estimand that the 0.5\% cluster design does not sign-identify.

\clearpage

\setcounter{table}{0}
\setcounter{figure}{0}

\section{Additional Figures and Tables}
\label{app:additional}

\begin{figure}[htpb!]
\caption{NSDAP Membership Before 1933}
\centering
\vspace{-.3cm}
\includegraphics[width=\textwidth]{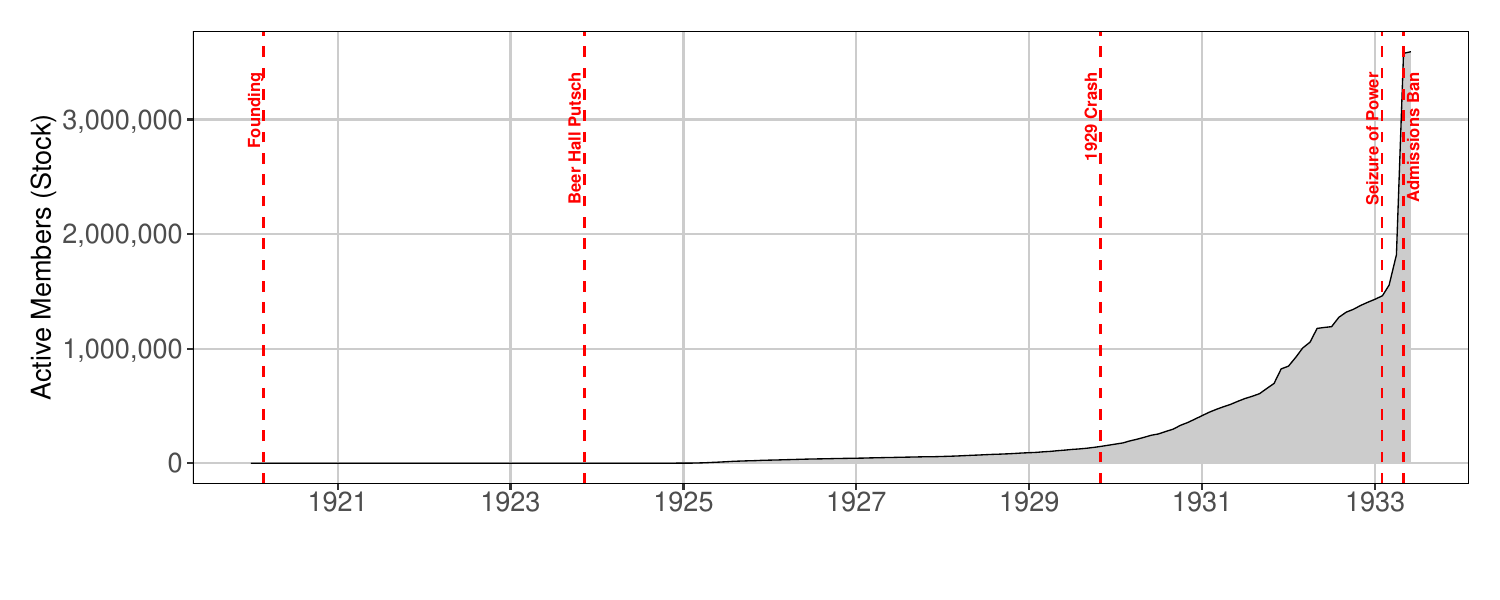}
\label{fig:membership_pre1933}
\\[4pt]
\justifying
\vspace{-1.35cm}
\scriptsize{\textbf{Note.} This figure shows NSDAP member counts before 1933.}
\end{figure}


\begin{figure}[htpb!]
\caption{Early Local Membership and Later Joining}
\centering
\includegraphics[width=0.85\textwidth]{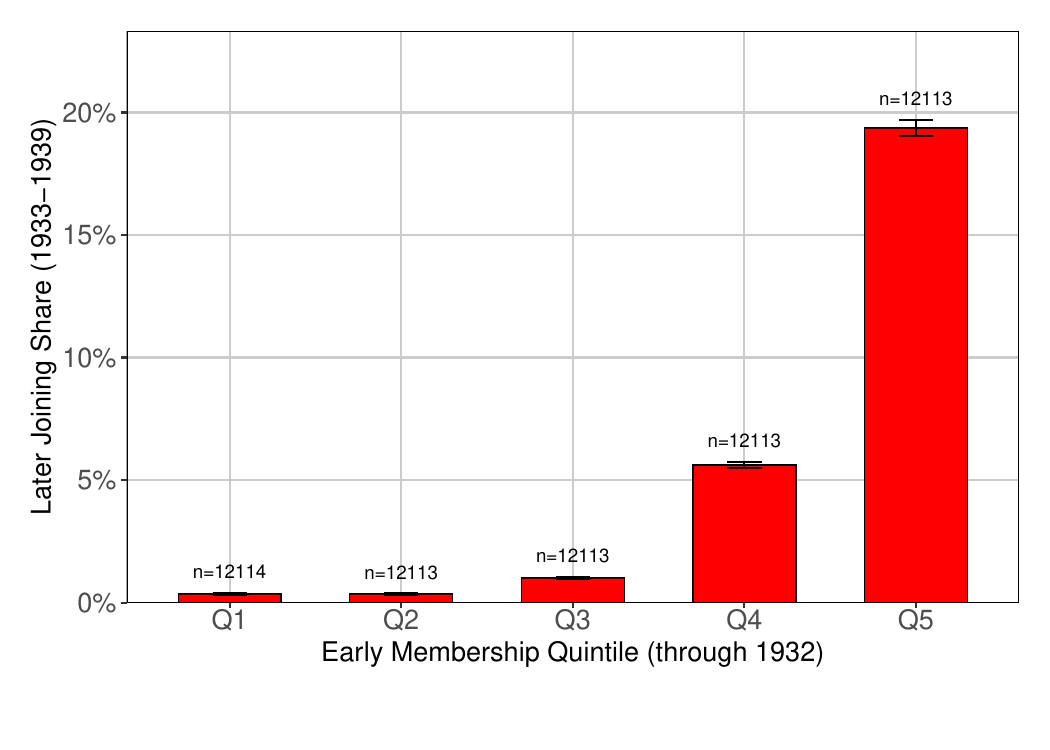}
\label{fig:early_later}
\\[4pt]
\justifying
\scriptsize{\textbf{Note.} Municipalities bucketed by cumulative NSDAP members through 1932. Bars show the mean share of the remaining 1933 population that joins during 1933--1939, with 95\% confidence intervals.}
\end{figure}

\begin{figure}[htpb!]
\caption{Cross-Wave Persistence of Municipal Mobilisation}
\centering
\includegraphics[width=0.75\textwidth]{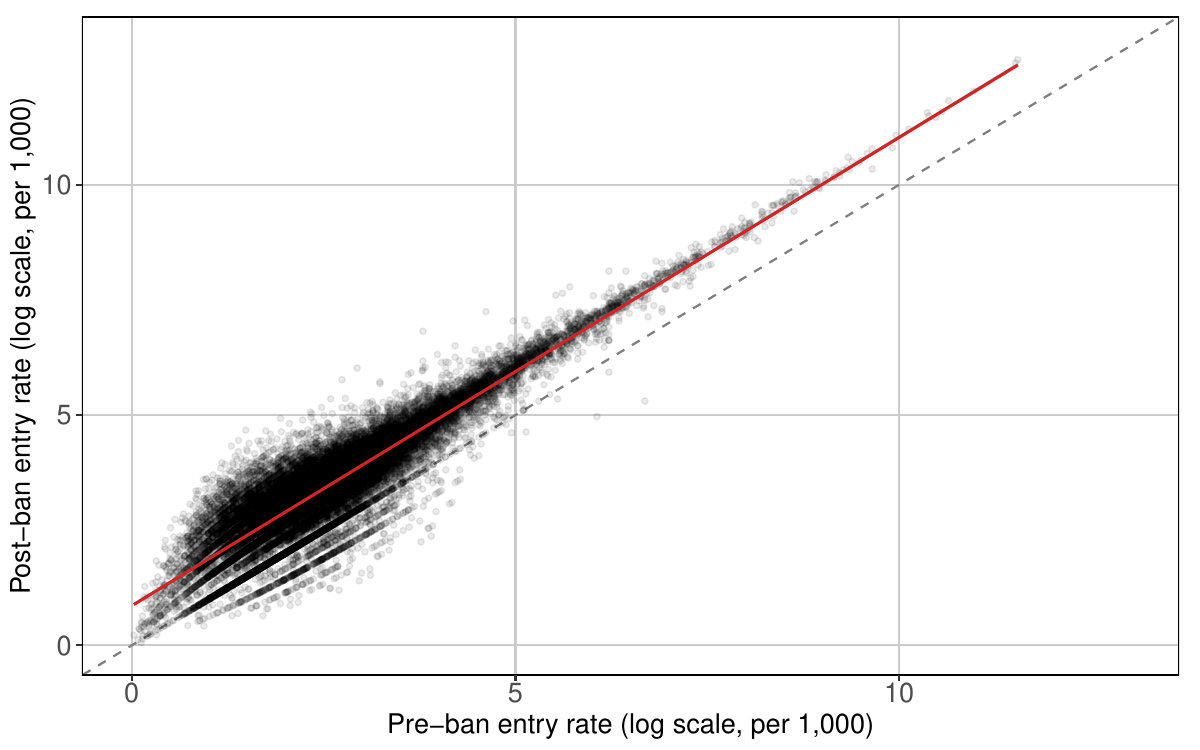}
\label{fig:cross_wave_persistence}
\\[4pt]
\justifying
\scriptsize{\textbf{Note.} Municipal NSDAP entry rates in the 12 months preceding the admissions ban (May~1932--April~1933, horizontal axis) and the 12 months following its lift (May~1937--April~1938, vertical axis), expressed per 1{,}000 residents and plotted on a $\log(1+x)$ scale. Each point is one of the $N=24{,}837$ municipalities with positive entry in both windows. The dashed grey line marks $y=x$; the red line is an OLS fit with a 95\% confidence band. The log-log slope is $1.02$, Pearson~$r=0.91$, and Spearman~$\rho=0.87$. Excluded from the plot: $31{,}681$ municipalities with zero entry in both windows, $7{,}358$ with zero pre-ban and positive post-ban entry, and $1{,}349$ with positive pre-ban and zero post-ban entry.}
\end{figure}

\begin{figure}[htpb!]
\caption{Flow Map of Municipal Entry Growth}
\centering
\includegraphics[width=0.8\textwidth]{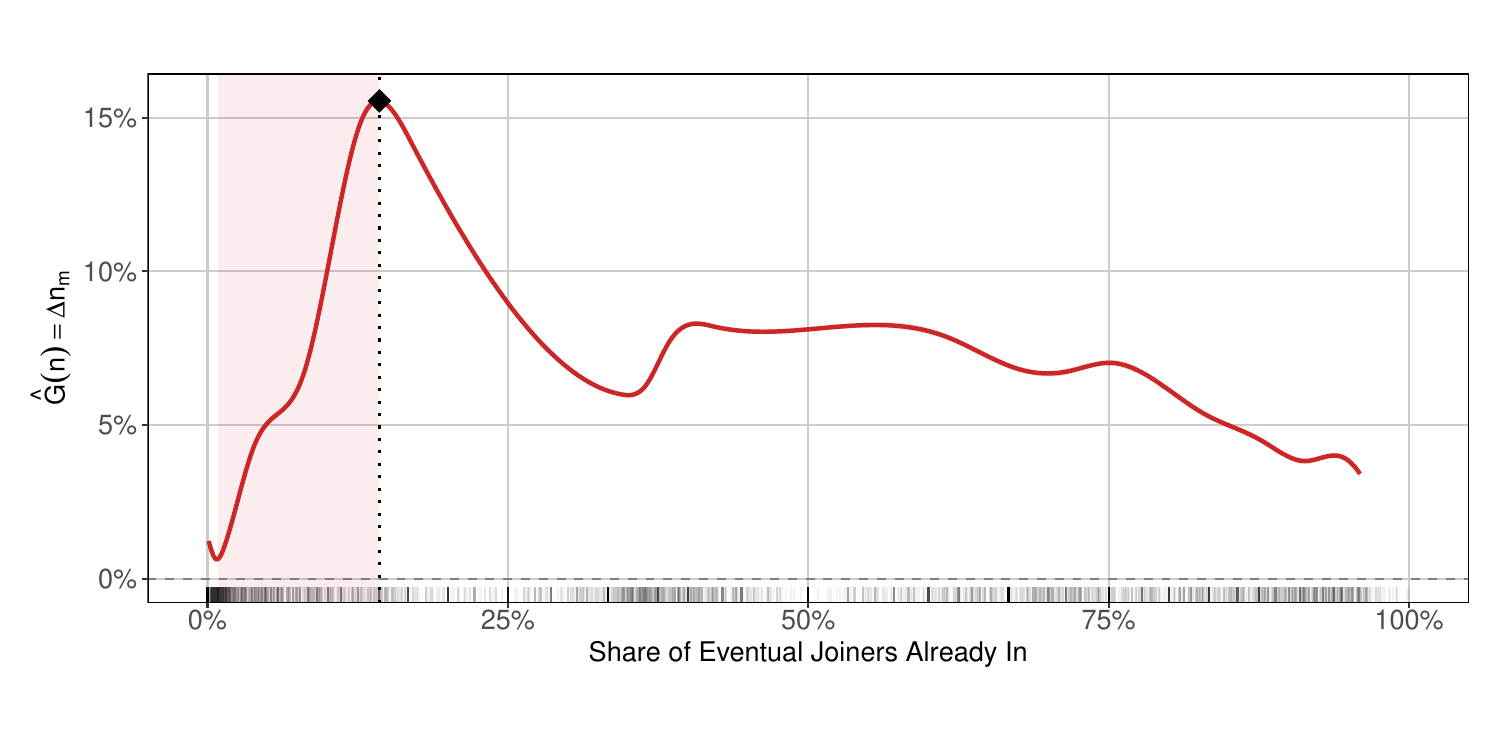}
\label{fig:flow_map_appendix}
\\[4pt]
\justifying
\scriptsize{\textbf{Note.} Descriptive flow map based on annual December snapshots. The horizontal axis is the lagged share of eventual municipal joiners who had already joined the NSDAP, and the vertical axis is the year-over-year change in that share. The curve is estimated using a generalized additive model with a cubic regression spline ($k=20$) and REML smoothing; the shaded region is a 95\% confidence band. The diamond marks the peak of the fitted curve and the rug marks show a random subsample of observed lagged shares. We read the subsequent flattening as descriptive evidence of saturation in the pool of eventual joiners, not as a separate causal test of coordination.}
\end{figure}

\begin{figure}[htpb!]
\caption{Community Composition: Unweighted-Benchmark Analogue of Figure~\ref{fig:community_butterfly}}
\centering
\includegraphics[width=\textwidth]{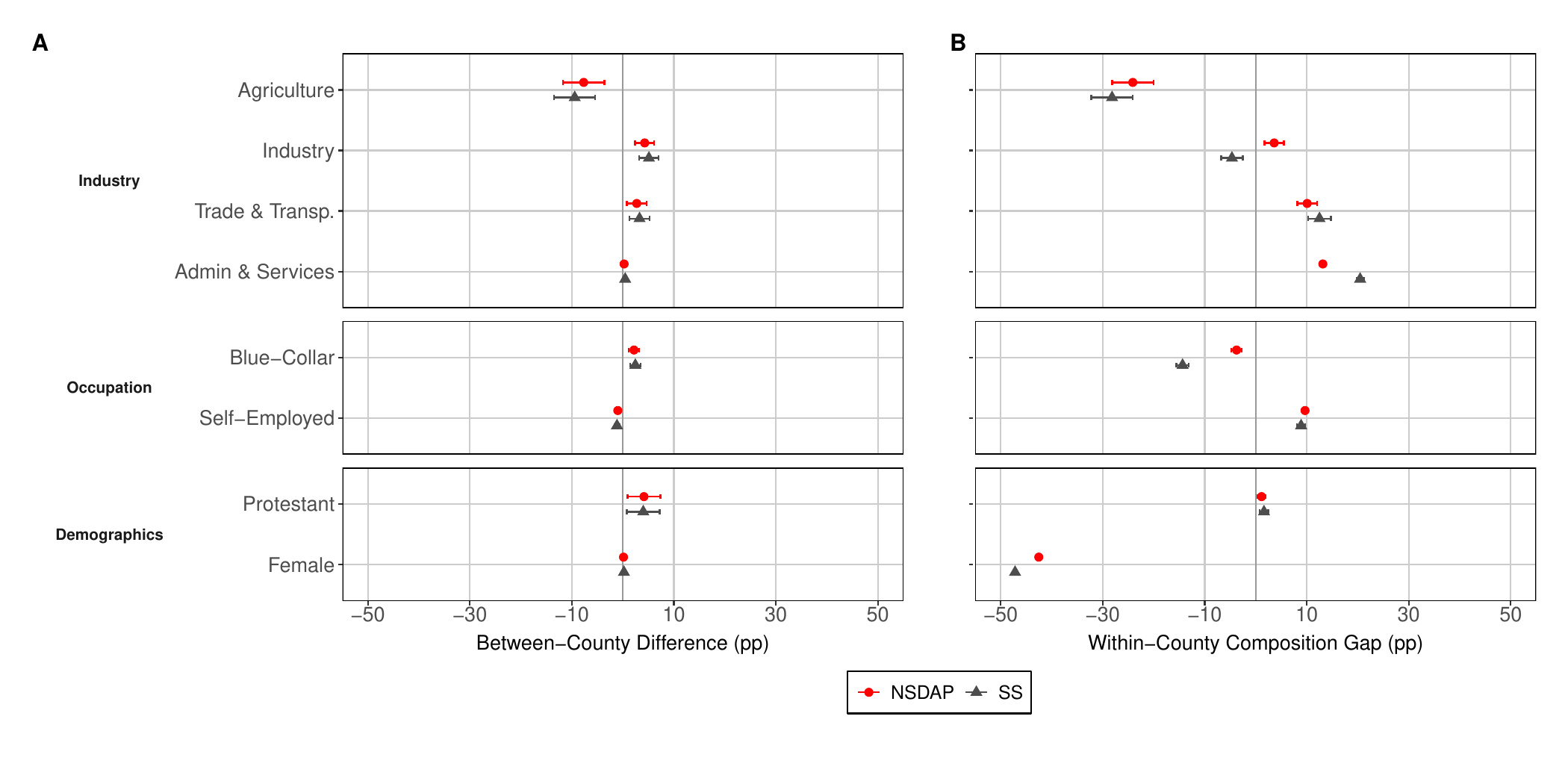}
\label{fig:community_butterfly_unweighted}
\\[4pt]
\justifying
\scriptsize{\textbf{Note.} Appendix companion to Figure~\ref{fig:community_butterfly}. Panel~\textbf{A} plots $\Delta_k^{g,\textup{between}}$ using an \emph{unweighted} cross-county mean $\bar{x}_k = C^{-1}\sum_c x_{ck}$ in place of the denominator-consistent population-weighted benchmark used in the main text. Panel~\textbf{B} is identical to the main-text Panel~B. NSDAP ($g=N$) in red, SS ($g=S$) in grey. Values in percentage points; error bars show 95\% confidence intervals with standard errors clustered at the county level. The unweighted benchmark treats every county equally regardless of population; the qualitative ordering of characteristics is preserved, but the Panel~A magnitudes differ because small counties receive disproportionate weight.}
\end{figure}

\begin{table}[hbt!]
\centering
\caption{The March 1933 Wave: Pre-March Membership Composition}
\label{table:gleich_composition}
\resizebox{\textwidth}{!}{%

\begingroup
\centering
\begin{tabular}{lccccccccc}
   \toprule
    & Female & Agric. & Admin/Svc & Trade & Industry & Protest. & Blue-Coll. & Self-Empl. & Joint \\ \cmidrule(lr){2-2} \cmidrule(lr){3-3} \cmidrule(lr){4-4} \cmidrule(lr){5-5} \cmidrule(lr){6-6} \cmidrule(lr){7-7} \cmidrule(lr){8-8} \cmidrule(lr){9-9} \cmidrule(lr){10-10}
    & \multicolumn{9}{c}{NSDAP membership share}\\
                                                   & (1)           & (2)           & (3)           & (4)           & (5)           & (6)           & (7)           & (8)           & (9)\\  
   \midrule 
   Post March 1933 x Any NSDAP Member              & 0.02$^{***}$  & 0.02$^{***}$  & 0.02$^{***}$  & 0.03$^{***}$  & 0.03$^{***}$  & 0.03$^{***}$  & 0.03$^{***}$  & 0.03$^{***}$  & 0.004$^{***}$\\   
                                                   & (0.0003)      & (0.0003)      & (0.0003)      & (0.0003)      & (0.0003)      & (0.0004)      & (0.0003)      & (0.0003)      & (0.0005)\\   
   Post March 1933 x Above med. Female             & 0.02$^{***}$  &               &               &               &               &               &               &               & 0.01$^{***}$\\   
                                                   & (0.0003)      &               &               &               &               &               &               &               & (0.0003)\\   
   Post March 1933 x Above med. Agriculture        &               & 0.01$^{***}$  &               &               &               &               &               &               & 0.01$^{***}$\\   
                                                   &               & (0.0004)      &               &               &               &               &               &               & (0.0004)\\   
   Post March 1933 x Above med. Admin \& Services  &               &               & 0.01$^{***}$  &               &               &               &               &               & 0.009$^{***}$\\   
                                                   &               &               & (0.0004)      &               &               &               &               &               & (0.0004)\\   
   Post March 1933 x Above med. Trade \& Transp.   &               &               &               & 0.007$^{***}$ &               &               &               &               & 0.006$^{***}$\\   
                                                   &               &               &               & (0.0004)      &               &               &               &               & (0.0004)\\   
   Post March 1933 x Above med. Industry           &               &               &               &               & 0.003$^{***}$ &               &               &               & 0.006$^{***}$\\   
                                                   &               &               &               &               & (0.0003)      &               &               &               & (0.0004)\\   
   Post March 1933 x Above med. Protestant         &               &               &               &               &               & 0.003$^{***}$ &               &               & 0.004$^{***}$\\   
                                                   &               &               &               &               &               & (0.0005)      &               &               & (0.0004)\\   
   Post March 1933 x Above med. Blue-Collar        &               &               &               &               &               &               & 0.002$^{***}$ &               & 0.002$^{***}$\\   
                                                   &               &               &               &               &               &               & (0.0003)      &               & (0.0004)\\   
   Post March 1933 x Above med. Self-Employed      &               &               &               &               &               &               &               & 0.005$^{***}$ & 0.0004\\   
                                                   &               &               &               &               &               &               &               & (0.0004)      & (0.0004)\\   
   Municipality FE                                 & \checkmark    & \checkmark    & \checkmark    & \checkmark    & \checkmark    & \checkmark    & \checkmark    & \checkmark    & \checkmark\\   
   Month FE                                        & \checkmark    & \checkmark    & \checkmark    & \checkmark    & \checkmark    & \checkmark    & \checkmark    & \checkmark    & \checkmark\\   
   Mean Outcome                                    & 0.0237        & 0.0237        & 0.0237        & 0.0237        & 0.0237        & 0.0237        & 0.0237        & 0.0237        & 0.0237\\  
    \\
   Observations                                    & 1,761,720     & 1,761,720     & 1,761,720     & 1,761,720     & 1,761,720     & 1,761,720     & 1,761,720     & 1,761,720     & 1,761,720\\  
   R$^2$                                           & 0.93492       & 0.93289       & 0.93215       & 0.93056       & 0.93005       & 0.93005       & 0.93003       & 0.93031       & 0.93810\\  
    \\
   \bottomrule
\end{tabular}
\par\endgroup

}
\\[4pt]
\justifying
\scriptsize{\textbf{Note.} Municipality-by-month panel, January~1932--June~1934. Dependent variable: municipal NSDAP membership share. $\text{Post March 1933}_t$ equals one from March~1933 onward. Columns~1--8 interact $\text{Post}$ with a binary indicator for whether municipality~$m$ lies above the cross-municipality median on the indicated share of its pre-March members, controlling for $\text{Post}\times\text{Any NSDAP Member}$. The Protestant share is the mean of the individual-level Bayesian name-based Protestant probability across pre-March members. Column~9 enters all eight interactions jointly. Municipality and month fixed effects included. Standard errors clustered at the county level in parentheses. Significance: $^{*}$~$p<0.10$, $^{**}$~$p<0.05$, $^{***}$~$p<0.01$.}
\end{table}

\begin{table}[hbt!]
\centering
\caption{The March 1933 Wave: Pre-March Paramilitary Presence}
\label{table:gleich_paramilitary}
\resizebox{.5\textwidth}{!}{%

\begingroup
\centering
\begin{tabular}{lcc}
   \toprule
    & SA & SS \\ \cmidrule(lr){2-2} \cmidrule(lr){3-3}
    & \multicolumn{2}{c}{NSDAP membership share}\\
                                      & (1)           & (2)\\  
   \midrule 
   Post March 1933 x Any NSDAP Member & 0.03$^{***}$  & 0.03$^{***}$\\   
                                      & (0.0003)      & (0.0003)\\   
   Post March 1933 x SA               & 0.002         &   \\   
                                      & (0.002)       &   \\   
   Post March 1933 x SS               &               & 0.006$^{***}$\\   
                                      &               & (0.0008)\\   
   Municipality FE                    & \checkmark    & \checkmark\\   
   Month FE                           & \checkmark    & \checkmark\\   
   Mean Outcome                       & 0.0237        & 0.0237\\  
    \\
   Observations                       & 1,761,720     & 1,761,720\\  
   R$^2$                              & 0.93277       & 0.93288\\  
    \\
   \bottomrule
\end{tabular}
\par\endgroup

}
\\[4pt]
\justifying
\scriptsize{\textbf{Note.} Municipality-by-month panel, January~1932--June~1934. Dependent variable: municipal NSDAP membership share. Column~1 interacts $\text{Post March 1933}_t$ with a binary indicator for whether municipality~$m$ had at least one SA member who joined before March~1933; column~2 does the same for the SS. Both columns control for $\text{Post}\times\text{Any NSDAP Member}$. SA and SS flags are constructed from the individual-level paramilitary indicator and from the \textit{ns\_organization} field on the membership card. Municipality and month fixed effects included. Standard errors clustered at the county level in parentheses. Significance: $^{*}$~$p<0.10$, $^{**}$~$p<0.05$, $^{***}$~$p<0.01$.}
\end{table}


\begin{figure}[htpb!]
\caption{NSDAP vs.\ SS: Individual-Level Characteristics}
\centering
\includegraphics[width=\textwidth]{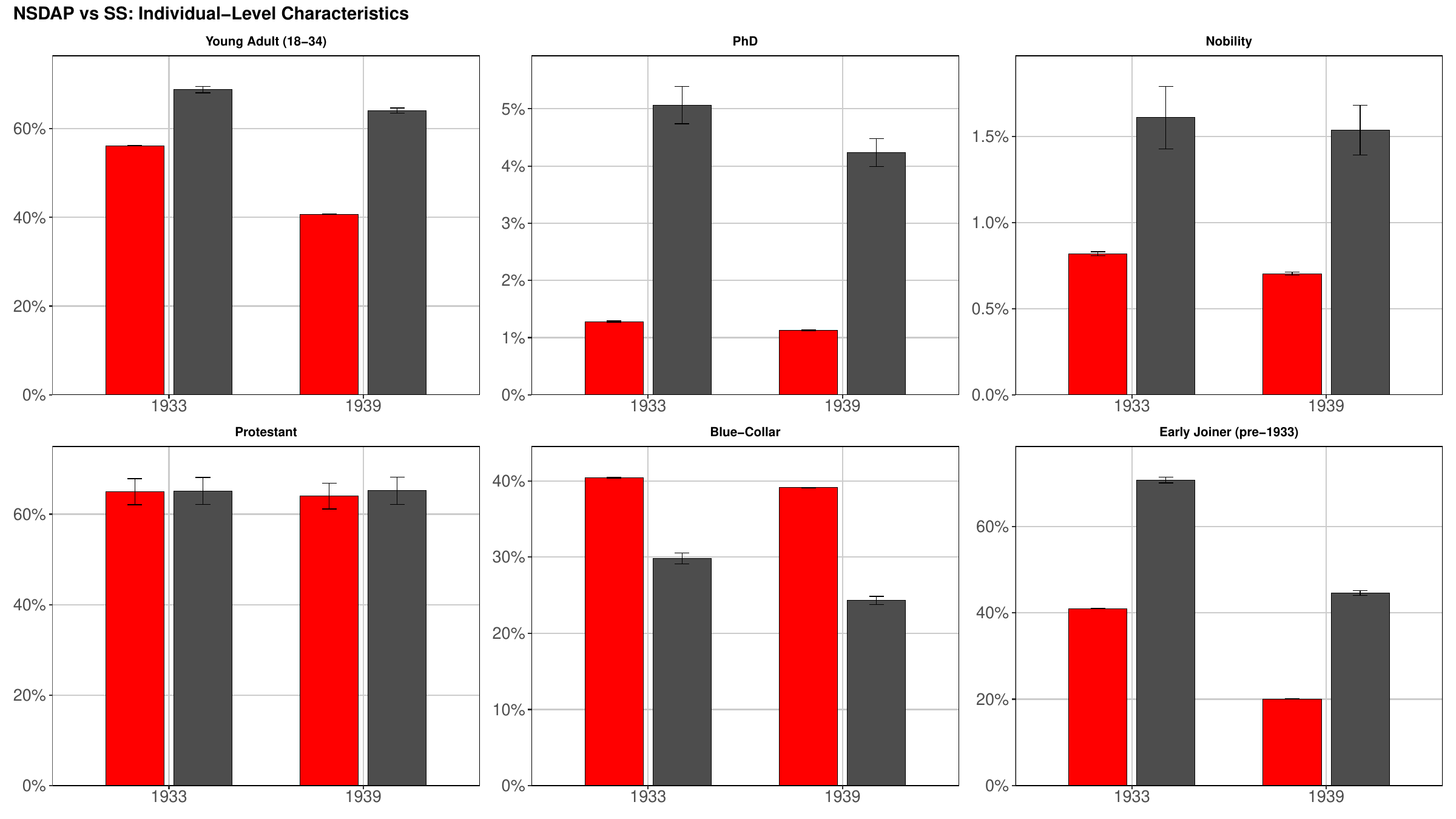}
\label{fig:nsdap_ss}
\\[4pt]
\justifying
\scriptsize{\textbf{Note.} County-level mean shares for NSDAP (red) and SS (dark grey) at 1933 and 1939. Six characteristics shown: Young Adult (18--34), PhD, Nobility, Protestant, Blue-Collar, and Early Joiner (pre-1933). Error bars show 95\% confidence intervals.}
\end{figure}

\clearpage

\clearpage
\printbibliography[title={Appendix References}]

\end{refsection}

\end{document}